\documentclass[longauth]{aa}
\usepackage{graphicx}
\usepackage[varg]{txfonts}
\usepackage{hyperref}
\usepackage{euclid}
\usepackage{bm}
\usepackage[dvipsnames]{xcolor}
\def\be{\begin{equation}}
\def\ee{\end{equation}}
\def\ba{\begin{eqnarray}}
\def\ea{\end{eqnarray}}
\def\eqi{\begin{equation}}
\def\eqf{\end{equation}}
\def\eqia{\begin{eqnarray}}
\def\eqfa{\end{eqnarray}}
\def\lcdm{$\Lambda$CDM }

\usepackage[nameinlink,noabbrev]{cleveref}
\Crefname{equation}{Eq.}{Eqs.}
\Crefname{eqnarray}{Eq.}{Eqs.}
\Crefname{section}{Sect.}{Sects.}
\Crefname{figure}{Fig.}{Figs.}
\crefname{equation}{Equation}{Equations}
\crefname{section}{Section}{Sections}
\crefname{figure}{Figure}{Figures}

\defcitealias{IST:paper1}{EC19}

\begin{document} 

\title{\Euclid: Forecast constraints on the cosmic distance duality relation with complementary external probes\thanks{This paper is published on behalf of the Euclid Consortium.}}

\titlerunning{\Euclid: Forecast constraints on the distance duality relation}

\author{M.~Martinelli$^{1}$\thanks{\email{matteo.martinelli@uam.es}}, C.J.A.P.~Martins$^{2,3}$, S.~Nesseris$^{1}$, D.~Sapone$^{4}$, I.~Tutusaus$^{5,6}$, A.~Avgoustidis$^{7}$, S.~Camera$^{8,9,10}$, C.~Carbone$^{11}$, S.~Casas$^{12}$, S.~Ili\'c$^{13,14,15}$, Z.~Sakr$^{15,16}$, V.~Yankelevich$^{17}$, N.~Auricchio$^{18}$, A.~Balestra$^{19}$, C.~Bodendorf$^{20}$, D.~Bonino$^{10}$, E.~Branchini$^{21,22,23}$, M.~Brescia$^{24}$, J.~Brinchmann$^{3}$, V.~Capobianco$^{10}$, J.~Carretero$^{25}$, M.~Castellano$^{23}$, S.~Cavuoti$^{24,26,27}$, R.~Cledassou$^{28}$, G.~Congedo$^{29}$, L.~Conversi$^{30,31}$, L.~Corcione$^{10}$, F.~Dubath$^{32}$, A.~Ealet$^{33}$, M.~Frailis$^{34}$, E.~Franceschi$^{18}$, M.~Fumana$^{11}$, B.~Garilli$^{11}$, B.~Gillis$^{29}$, C.~Giocoli$^{35,36,37}$, F.~Grupp$^{20,38}$, S.V.H.~Haugan$^{39}$, W.~Holmes$^{40}$, F.~Hormuth$^{41}$, K.~Jahnke$^{42}$, S.~Kermiche$^{43}$, M.~Kilbinger$^{12}$, T.D.~Kitching$^{44}$, B.~Kubik$^{45}$, M.~Kunz$^{46}$, H.~Kurki-Suonio$^{47}$, S.~Ligori$^{10}$, P.B.~Lilje$^{39}$, I.~Lloro$^{48}$, O.~Marggraf$^{49}$, K.~Markovic$^{40}$, R.~Massey$^{50}$, S.~Mei$^{51}$, M.~Meneghetti$^{35,37}$, G.~Meylan$^{52}$, L.~Moscardini$^{18,36,53}$, S.~Niemi$^{44}$, C.~Padilla$^{25}$, S.~Paltani$^{32}$, F.~Pasian$^{34}$, V.~Pettorino$^{12}$, S.~Pires$^{12}$, G.~Polenta$^{54}$, M.~Poncet$^{28}$, L.~Popa$^{55}$, L.~Pozzetti$^{18}$, F.~Raison$^{20}$, J.~Rhodes$^{40}$, M.~Roncarelli$^{18,36}$, R.~Saglia$^{20,38}$, P.~Schneider$^{49}$, A.~Secroun$^{43}$, S.~Serrano$^{5,6}$, C.~Sirignano$^{56,57}$, G.~Sirri$^{53}$, F.~Sureau$^{12}$, A.N.~Taylor$^{29}$, I.~Tereno$^{58,59}$, R.~Toledo-Moreo$^{60}$, L.~Valenziano$^{18,53}$, T.~Vassallo$^{38}$, Y.~Wang$^{61}$, N.~Welikala$^{29}$, J.~Weller$^{20,38}$, A.~Zacchei$^{34}$}

\institute{$^{1}$ Instituto de F\'isica T\'eorica UAM-CSIC, Campus de Cantoblanco, E-28049 Madrid, Spain\\
$^{2}$ Centro de Astrof\'{\i}sica da Universidade do Porto, Rua das Estrelas, 4150-762 Porto, Portugal\\
$^{3}$ Instituto de Astrof\'isica e Ci\^encias do Espa\c{c}o, Universidade do Porto, CAUP, Rua das Estrelas, PT4150-762 Porto, Portugal\\
$^{4}$ Departamento de F\'isica, FCFM, Universidad de Chile, Blanco Encalada 2008, Santiago, Chile\\
$^{5}$ Institute of Space Sciences (ICE, CSIC), Campus UAB, Carrer de Can Magrans, s/n, 08193 Barcelona, Spain\\
$^{6}$ Institut d’Estudis Espacials de Catalunya (IEEC), 08034 Barcelona, Spain\\
$^{7}$ School of Physics and Astronomy, University of Nottingham, University Park, Nottingham NG7 2RD, UK\\
$^{8}$ INFN-Sezione di Torino, Via P. Giuria 1, I-10125 Torino, Italy\\
$^{9}$ Dipartimento di Fisica, Universit\'a degli Studi di Torino, Via P. Giuria 1, I-10125 Torino, Italy\\
$^{10}$ INAF-Osservatorio Astrofisico di Torino, Via Osservatorio 20, I-10025 Pino Torinese (TO), Italy\\
$^{11}$ INAF-IASF Milano, Via Alfonso Corti 12, I-20133 Milano, Italy\\
$^{12}$ AIM, CEA, CNRS, Universit\'e Paris-Saclay, Universit\'{e} Paris Diderot, Sorbonne Paris Cit\'{e}, F-91191 Gif-sur-Yvette, France\\
$^{13}$ Universit\'e PSL, Observatoire de Paris, Sorbonne Universit\'e, CNRS, LERMA, F-75014, Paris, France\\
$^{14}$ CEICO, Institute of Physics of the Czech Academy of Sciences, Na Slovance 2, Praha 8, Czech Republic\\
$^{15}$ Institut de Recherche en Astrophysique et Plan\'etologie (IRAP), Universit\'e de Toulouse, CNRS, UPS, CNES, 14 Av. Edouard Belin, F-31400 Toulouse, France\\
$^{16}$ Universit\'e St Joseph; UR EGFEM, Faculty of Sciences, Beirut, Lebanon\\
$^{17}$ Astrophysics Research Institute, Liverpool John Moores University, 146 Brownlow Hill, Liverpool L3 5RF, UK\\
$^{18}$ INAF-Osservatorio di Astrofisica e Scienza dello Spazio di Bologna, Via Piero Gobetti 93/3, I-40129 Bologna, Italy\\
$^{19}$ INAF-Osservatorio Astronomico di Padova, Via dell'Osservatorio 5, I-35122 Padova, Italy\\
$^{20}$ Max Planck Institute for Extraterrestrial Physics, Giessenbachstr. 1, D-85748 Garching, Germany\\
$^{21}$ INFN-Sezione di Roma Tre, Via della Vasca Navale 84, I-00146, Roma, Italy\\
$^{22}$ Department of Mathematics and Physics, Roma Tre University, Via della Vasca Navale 84, I-00146 Rome, Italy\\
$^{23}$ INAF-Osservatorio Astronomico di Roma, Via Frascati 33, I-00078 Monteporzio Catone, Italy\\
$^{24}$ INAF-Osservatorio Astronomico di Capodimonte, Via Moiariello 16, I-80131 Napoli, Italy\\
$^{25}$ Institut de F\'{i}sica d’Altes Energies (IFAE), The Barcelona Institute of Science and Technology, Campus UAB, 08193 Bellaterra (Barcelona), Spain\\
$^{26}$ Department of Physics "E. Pancini", University Federico II, Via Cinthia 6, I-80126, Napoli, Italy\\
$^{27}$ INFN section of Naples, Via Cinthia 6, I-80126, Napoli, Italy\\
$^{28}$ Centre National d'Etudes Spatiales, Toulouse, France\\
$^{29}$ Institute for Astronomy, University of Edinburgh, Royal Observatory, Blackford Hill, Edinburgh EH9 3HJ, UK\\
$^{30}$ European Space Agency/ESRIN, Largo Galileo Galilei 1, 00044 Frascati, Roma, Italy\\
$^{31}$ ESAC/ESA, Camino Bajo del Castillo, s/n., Urb. Villafranca del Castillo, 28692 Villanueva de la Ca\~nada, Madrid, Spain\\
$^{32}$ Department of Astronomy, University of Geneva, ch. d'\'Ecogia 16, CH-1290 Versoix, Switzerland\\
$^{33}$ Univ Lyon, Univ Claude Bernard Lyon 1, CNRS/IN2P3, IP2I Lyon, UMR 5822, F-69622, Villeurbanne, France\\
$^{34}$ INAF-Osservatorio Astronomico di Trieste, Via G. B. Tiepolo 11, I-34131 Trieste, Italy\\
$^{35}$ Istituto Nazionale di Astrofisica (INAF) - Osservatorio di Astrofisica e Scienza dello Spazio (OAS), Via Gobetti 93/3, I-40127 Bologna, Italy\\
$^{36}$ Dipartimento di Fisica e Astronomia, Universit\'a di Bologna, Via Gobetti 93/2, I-40129 Bologna, Italy\\
$^{37}$ Istituto Nazionale di Fisica Nucleare, Sezione di Bologna, Via Irnerio 46, I-40126 Bologna, Italy\\
$^{38}$ Universit\"ats-Sternwarte M\"unchen, Fakult\"at f\"ur Physik, Ludwig-Maximilians-Universit\"at M\"unchen, Scheinerstrasse 1, 81679 M\"unchen, Germany\\
$^{39}$ Institute of Theoretical Astrophysics, University of Oslo, P.O. Box 1029 Blindern, N-0315 Oslo, Norway\\
$^{40}$ Jet Propulsion Laboratory, California Institute of Technology, 4800 Oak Grove Drive, Pasadena, CA, 91109, USA\\
$^{41}$ von Hoerner \& Sulger GmbH, Schlo{\ss}Platz 8, D-68723 Schwetzingen, Germany\\
$^{42}$ Max-Planck-Institut f\"ur Astronomie, K\"onigstuhl 17, D-69117 Heidelberg, Germany\\
$^{43}$ Aix-Marseille Univ, CNRS/IN2P3, CPPM, Marseille, France\\
$^{44}$ Mullard Space Science Laboratory, University College London, Holmbury St Mary, Dorking, Surrey RH5 6NT, UK\\
$^{45}$ University of Lyon, UCB Lyon 1, CNRS/IN2P3, IUF, IP2I Lyon, France\\
$^{46}$ Universit\'e de Gen\`eve, D\'epartement de Physique Th\'eorique and Centre for Astroparticle Physics, 24 quai Ernest-Ansermet, CH-1211 Gen\`eve 4, Switzerland\\
$^{47}$ Department of Physics and Helsinki Institute of Physics, Gustaf H\"allstr\"omin katu 2, 00014 University of Helsinki, Finland\\
$^{48}$ NOVA optical infrared instrumentation group at ASTRON, Oude Hoogeveensedijk 4, 7991PD, Dwingeloo, The Netherlands\\
$^{49}$ Argelander-Institut f\"ur Astronomie, Universit\"at Bonn, Auf dem H\"ugel 71, 53121 Bonn, Germany\\
$^{50}$ Institute for Computational Cosmology, Department of Physics, Durham University, South Road, Durham, DH1 3LE, UK\\
$^{51}$ Universit\'e de Paris, F-75013, Paris, France, LERMA, Observatoire de Paris, PSL Research University, CNRS, Sorbonne Universit\'e, F-75014 Paris, France\\
$^{52}$ Observatoire de Sauverny, Ecole Polytechnique F\'ed\'erale de Lau- sanne, CH-1290 Versoix, Switzerland\\
$^{53}$ INFN-Sezione di Bologna, Viale Berti Pichat 6/2, I-40127 Bologna, Italy\\
$^{54}$ Space Science Data Center, Italian Space Agency, via del Politecnico snc, 00133 Roma, Italy\\
$^{55}$ Institute of Space Science, Bucharest, Ro-077125, Romania\\
$^{56}$ INFN-Padova, Via Marzolo 8, I-35131 Padova, Italy\\
$^{57}$ Dipartimento di Fisica e Astronomia “G.Galilei", Universit\'a di Padova, Via Marzolo 8, I-35131 Padova, Italy\\
$^{58}$ Instituto de Astrof\'isica e Ci\^encias do Espa\c{c}o, Faculdade de Ci\^encias, Universidade de Lisboa, Tapada da Ajuda, PT-1349-018 Lisboa, Portugal\\
$^{59}$ Departamento de F\'isica, Faculdade de Ci\^encias, Universidade de Lisboa, Edif\'icio C8, Campo Grande, PT1749-016 Lisboa, Portugal\\
$^{60}$ Universidad Polit\'ecnica de Cartagena, Departamento de Electr\'onica y Tecnolog\'ia de Computadoras, 30202 Cartagena, Spain\\
$^{61}$ Infrared Processing and Analysis Center, California Institute of Technology, Pasadena, CA 91125, USA\\
}

\authorrunning{M. Martinelli et al.}

 
\abstract{
{In metric theories of gravity with photon number conservation, the luminosity and angular diameter distances are related via the Etherington relation, also known as the distance duality relation (DDR). A violation of this relation would rule out the standard cosmological paradigm and point to the presence of new physics.}
{We quantify the ability of \Euclid, in combination with contemporary surveys, to improve the current constraints on deviations from the DDR in the redshift range $0<z<1.6$.}
{We start with an analysis of the latest available data, improving previously reported constraints by a factor of 2.5. We then present a detailed analysis of simulated \Euclid and external data products, using both standard parametric methods (relying on phenomenological descriptions of possible DDR violations) and a machine learning reconstruction using genetic algorithms.}
{We find that for parametric methods \Euclid can (in combination with external probes) improve current constraints by approximately a factor of six, while for non-parametric methods \Euclid can improve current constraints by a factor of three.}
{Our results highlight the importance of surveys like \Euclid in accurately testing the pillars of the current cosmological paradigm and constraining physics beyond the standard cosmological model.}}


\maketitle
%

\section{Introduction \label{sec:intro}}

Standard cosmological analyses rely on several explicit or implicit assumptions. Three examples of commonly made assumptions are that the Copernican principle holds (i.e. we are not at a special place in the Universe), that the photon number is conserved, and that the Universe is homogeneous and isotropic, at least on sufficiently large scales. While they are, in some sense, the pillars on which standard cosmology is built and none of them are seriously challenged by current data \citep[for a possible exception see][]{Webb}, they are violated in many extensions of the standard cosmological model and of the standard particle physics paradigm, with extensions of the latter collectively known as beyond the standard model (BSM) theories. In this work, we focus on another of these assumptions, which is related to, but conceptually different from, the previous three: the so-called distance duality relation (DDR), which is crucial for tests of the background expansion rate of the Universe as it allows us to relate the luminosity and angular diameter distances while at the same time affecting the prediction for the change in redshift of the cosmic microwave background (CMB) radiation temperature. Specifically, the DDR holds true for metric theories of gravity where the photon number is conserved and photons travel along null geodesics. Testing these predictions with cosmological data therefore has the potential to rule out large classes of extended theories or to observe signatures of non-standard physics.

Combining supernova data from the supernova cosmology project (SCP) Union 2008 compilation \citep{Union08} and $H(z)$ data from \citet{SJVKS}, deviations from the standard DDR have been constrained to few percent \citep{Avgoustidis2010,Ma:2016bjt}. At the same time, direct measurements of CMB temperature at different redshifts, denoted $T(z)$ hereafter, have been obtained at both low redshifts ($z \lesssim 1$) via observations of the Sunyaev--Zel'dovich effect in galaxy clusters \citep{Luzzi2009} and at higher redshifts ($z > 1$) through high-resolution spectroscopy of atomic, ionic, or molecular levels excited by the absorption of  CMB photons \citep{Noterdaeme2010}. The deviation from the standard redshift evolution of CMB temperature has also been  constrained to a few percent \citep{Noterdaeme2010}. In models where deviations from the DDR arise due to an effective opacity effect (e.g. dimming due to photons decaying into an unobserved particle), the same physical mechanism can also give rise to a related violation of the standard $T(z)$ evolution. For such models, the combination of distance and temperature measurements as two independent probes of the same underlying mechanism shrunk the constraints on the deviation of $T(z)$ to $0.8\%$ \citep{Avgoustidis2015}. Other possible observables to constrain possible violations of the DDR include galaxy clusters \citep{Holanda:2010vb,Li:2011exa}, the Sunyaev-Zeldovich effect \citep{Holanda:2012at}, strong gravitational lensing \citep{Liao:2015uzb}, and standard sirens from gravitational wave observations \citep{Liao:2019xug,1809125}.

Upcoming and more sensitive cosmological surveys for supernovae and baryon acoustic oscillations (BAO) will further tighten the constraints achievable through distance measurements. Here we focus on \Euclid, an M-class space mission of the European Space Agency due for launch in 2022. It will carry two different instruments on board: a visible imager \citep{VIS_paper} and a near-infrared spectrophotometric instrument \citep{NISP_paper}. Together they will carry out a photometric and spectroscopic galaxy survey over 15\,000 deg$^2$ of extra-galactic sky with the aim of measuirng the geometry of the Universe and the growth of structures up to $z\sim 2$ and beyond~\citep{Laureijs:2011gra}.

\Euclid will have three main cosmological probes: weak lensing and galaxy clustering from the photometric survey and galaxy clustering from the spectroscopic survey. While photometric galaxy surveys allow for observations of large numbers of galaxies with relatively large redshift uncertainties, spectroscopic galaxy surveys provide information for fewer objects but with much higher radial precision.  The spectroscopic accuracy of \Euclid will allow for precise galaxy clustering analyses that include the radial dimension. Here we simulate BAO data from \Euclid using the Fisher matrix technique and specifically following the same strategy used in \citet[][]{IST:paper1}, hereafter EC19, for the spectroscopic survey.

On the other hand, improved direct measurements of the CMB temperature at different redshifts will be available, such as those expected in the coming years from ESPRESSO \citep{ESPRESSO1,ESPRESSO2}, and eventually ELT-HIRES \citep{HIRES}, and these will significantly improve the available constraining power \citep{Avgoustidis2013}. In addition to this, future observations of gravitational wave events will allow us to exploit standard sirens to obtain luminosity distance measurements at even higher redshifts, thus extending the redshift range of DDR tests \citep[see for example][]{Yang:2017bkv}.

Our work also highlights some of the synergies between different surveys. Specifically, we will show how data from the Dark Energy Spectroscopic Instrument \citep[DESI,][]{DESI2016}, a survey that aims at probing the expansion rate and large-scale structure (LSS) of the universe, and from the Legacy Survey of Space and Time (LSST), performed by the Vera C. Rubin Observatory \citep{Abell:2009aa}, can complement the \Euclid BAO survey and extend the probed redshift range.

In \citet{ReviewDoc}, forecast constraints on deviations from the standard DDR were presented (and their implications on mechanisms of cosmic opacity explored), combining a dark energy task force stage IV supernova mission \citep[][taking SNAP as a concrete example]{DEtaskforce} and the galaxy survey expected to be performed by the \Euclid satellite \citep{Laureijs:2011gra}. In this paper, we aim to update and extend those results on constraining DDR deviations; specifically, we rely on more recent \Euclid specifications \citepalias[see][]{IST:paper1} and we investigate possible synergies between this survey and contemporary observations. We also refine the analysis done in \citet{ReviewDoc}; we follow the common approach of encoding DDR violations in the phenomenological function $\epsilon(z)$, but, alongside the common constant parameterization, we include here a binning in redshift of this function in order to understand if current or future data are able to detect a redshift trend. Moreover, we also apply a more refined machine learning technique to reconstruct $\epsilon(z)$ with a minimal set of assumptions, performing our analysis with the use of genetic algorithms (GAs) \citep{Bogdanos:2009ib,Nesseris:2012tt}. 

After reviewing the theoretical background of the DDR in \Cref{sec:theory}, we describe the different analyses done in this work to test the relation in \Cref{sec:analysis}, detailing both our parameterized approach and the agnostic reconstruction. We then present the constraints obtained from current observations in \Cref{sec:curr_data}. The results of this analysis are used as a fiducial cosmology for the mock data we produce in \Cref{sec:fore_data} and then compared with forecast results, which are discussed in \Cref{sec:fore_res}. Finally, we draw our conclusions in \Cref{sec:conclusions}.

\section{Extensions of the distance duality relation}\label{sec:theory}

In any cosmological model based on a metric theory of gravity, the Etherington relation \citep{Etherington}, also known as the DDR,  
implies that distance measures are unique. The luminosity distance, $d_{\rm L}(z)$, is related to the angular diameter distance, $d_{\rm A}(z)$, as
\begin{equation}\label{eq:std_DDR}
    d_{\rm L}(z) = (1+z)^2d_{\rm A}(z)\, ,
\end{equation}
and this relation is valid in any cosmological background where photons travel on null geodesics and where, crucially, the photon number is conserved.
Deviations from DDR as expressed in Eq.~(\ref{eq:std_DDR}) can be directly constrained by data (for example using a phenomenological parameterization), but one is often interested in constraining the underlying physical mechanism giving rise to such deviations within a class of cosmological models. For example, in models where photons can decay into another particle that remains unobserved, there is an effective violation of photon number conservation which can be used to constrain the coupling between the photon and the new particle.

On the other hand, if the expansion of the Universe is adiabatic and the CMB spectrum was a black-body at the time it originated, such a property will be preserved by the subsequent cosmological evolution,  with the CMB temperature evolving as
\begin{equation}\label{eq:std_temp}
    T(z) = T_0 (1 + z)\, .
\end{equation} 
This is a robust prediction of standard cosmology, but it is violated in many non-standard models, including scenarios involving photon mixing \citep[for a review see][]{WISPs} and the violation of photon number conservation, which may also induce deviations from the DDR. Throughout this paper we assume that the cosmological principle holds, namely that the Universe is, to first approximation, homogeneous and isotropic. Consequently, we do not investigate any possible dependence of the DDR on the direction of the sky and only tackle its possible redshift dependence \citep[for further discussion on the validity of such an assumption see for example][]{Maartens:2011yx,Ntelis:2017nrj}.

From a theoretical point of view, models violating the DDR through a violation of photon number conservation are of particular interest and, as has been alluded to above, in such extended theories one expects deviations from both \Cref{eq:std_DDR} and \Cref{eq:std_temp}. Therefore, an analysis of such behaviour could in principle benefit from the complementarity between galaxy and supernova surveys, probing departures from the DDR, and spectroscopic tests of $T(z)$ evolution, both of which will be available during the next decade.

Deviations from the Etherington relation are commonly parameterized as
\begin{equation}\label{eq:epsz}
    d_{\rm L}(z)=(1+z)^{2+\epsilon(z)}d_{\rm A}(z)\,,
\end{equation}
where the function $\epsilon(z)$ is usually assumed to be constant and, using currently available data, its value is constrained to be $\mathcal{O}(10^{-2})$ \citep{Avgoustidis2009,Avgoustidis2010}.
As the precision of the data improves and the available redshift range is extended, the DDR could be probed at larger redshifts, $z\gtrsim 2$, and tighter constraints of a possible redshift dependence could, in principle, be obtained. Deviations from the standard DDR are also commonly encoded in a function $\eta(z)$, defined as
\begin{equation}\label{eq:eta1}
\eta(z) = \frac{d_{\rm L}(z)}{d_{\rm A}(z)(1 + z)^2} = (1 + z)^{\epsilon(z)}.
\end{equation}

Similarly, deviations from the standard evolution of the CMB temperature with redshift can be parametrized phenomenologically by \citep{2000MNRAS.312..747L,Luzzi2009,Avgoustidis2011}
\begin{equation}\label{symbol:Tzbeta}
T(z) = T_0 (1+z)^{1-\beta}\, ,
\end{equation}
where for simplicity it is assumed that violations of the standard behaviour are achromatic (they do not depend on the photons' wavelength) and approximately adiabatic, so the spectrum of CMB radiation remains approximately a black body spectrum. A discussion of these assumptions can be found in \citet{Avgoustidis2015}. It is important to stress that in a broad range of models where the photon number is not conserved, and so the temperature-redshift relation and the DDR are both violated, the functions parameterizing these two possible violations will not be independent. Defining a generic function $f(z)$ encoding the violation of the temperature-redshift relation as
\begin{equation}
T(z) = T_0 (1 + z) f(z)\, ,
\end{equation}
it is possible to show \citep{Avgoustidis2011} that the DDR violation will then be
\begin{equation}
d_{\rm L}(z) = d_{\rm A}(z)(1 + z)^2 f(z)^{3/2}\,.
\end{equation}

Therefore, for the two simple parameterizations introduced above, the parameterized deviations from the standard model are related as 
\begin{equation}
    \epsilon= - \frac{3}{2}\beta\, .
\end{equation}

In this work we focus mainly on constraining violations from the DDR using supernova and BAO data, but have in mind a specific class of physical mechanisms producing violations of both \Cref{eq:std_DDR} and \Cref{eq:std_temp}, due to a change in the photon flux during the propagation from distant sources. Such mechanisms would affect the supernova luminosity distance measures but not the determinations of the angular diameter distance. This means that probes of the latter (BAO) can be combined with supernova surveys to constrain deviations from photon number conservation. In addition, it also means that direct measurements of the CMB temperature at different redshifts could be used to further improve the available constraining power, as mentioned above.

Photon conservation can be violated by simple astrophysical effects or by exotic physics. Amongst the former we find, for instance, attenuation due to interstellar dust, gas, and/or plasma. Such astrophysical mechanisms produce an effective opacity, which would correspond to a positive value of the phenomenological parameter $\epsilon$. Most known sources of attenuation are expected to be clustered and can be typically constrained down to the $0.1\%$ level \citep{Menard2008,More2009}.

Unclustered sources of attenuation are more difficult to constrain. Grey dust \citep{Aguirre1999} was initially invoked to explain the observed dimming of Type~Ia supernovae (SnIa) without resorting to cosmic acceleration. While this has been subsequently ruled out by observations \citep{Aguirre1999b,BassettKunz2004b}, it has been shown \citep{Corasaniti2006} that the effect of grey dust could cause an extinction as large as 0.08 mag at $z = 1.7$, thus potentially affecting dark energy parameter inference from future supernova surveys.

Concerning exotic physics explanations, a possible source of photon conservation violation is the coupling of photons with particles beyond the standard model of particle physics. Such couplings would mean that, while passing through the intergalactic medium, a photon could disappear, or even (re)appear, while interacting  with such exotic particles, modifying the apparent luminosity of sources. Therefore, in this case, we may in principle envisage both positive and negative values for $\epsilon$. In \citet{Avgoustidis2010}, the mixing of photons with several such particles is considered and constrained in three representative scenarios: scalars known as axion-like particles \citep{SvrcekWitten2006}, chameleons \citep{Brax2010}, and the possibility of mini-charged particles, which have a tiny and unquantized electric charge \citep{Holdom1986,BatellGherghetta2006}. The implications of each of these three specific scenarios for the SnIa luminosity have been described by several authors \citep{Csaki2002, Mortsell2002,Burrage2008,Ahlers2009}.

Finally, it is worth noting that any violations in photon conservation can be described as an opacity effect in the observed luminosity distance, which one can parameterize through a generic opacity parameter, $\tau(z)$, as
\begin{equation}
d_{\rm L,\mathrm{obs}}^2=d^2_{\rm L,\mathrm{true}}\exp[\tau(z)]\,.
\end{equation}

We note that a negative $\tau(z)$ allows for apparent brightening of light sources, as would be the case, for example, if exotic particles were also emitted from the source and converted
into photons along the line of sight \citep[see][]{Burrage2008}. For specific models of exotic matter-photon coupling, such as axion-like  particles, chameleons, and mini-charged particles, the function $\tau(z)$ can be obtained in terms of the parameters of the model \citep{Avgoustidis2010}.

\section{Analysis method}\label{sec:analysis}

In this paper, we aim at obtaining constraints on possible deviations from the standard DDR, without assuming any specific model, from both current and mock data. For this reason we adopt two different approaches: on the one hand we parameterize the $\epsilon(z)$ function, both as a constant and binning it in redshift, while on the other hand we also adopt a more general approach based on machine learning, reconstructing the function with GAs. In this section we review in detail the two approaches.

\subsection{Parameterized approach}\label{sec:parapp}
A first simple way to constrain the cosmic DDR is to parameterize departures from the Etherington relation through a constant (redshift-independent) parameter $\epsilon_0$, that is
\begin{equation}\label{eq:eps0}
    d_{\rm L}(z) = (1+z)^{2+\epsilon_0}d_{\rm A}(z)\,,
\end{equation}
with $\epsilon_0=0$ being the standard limit. However, we are also interested in a possible redshift dependence of such departures, as many of the theoretical models discussed in \Cref{sec:theory} produce a redshift dependent modification of the DDR. Therefore, we take one step further by using the general form of \Cref{eq:epsz}. Choosing a specific model violating the Etherington relation would allow us to obtain $\epsilon(z)$ in terms of the parameters of the chosen model.
However, the aim of this paper is not to constrain specific theories; in order not to make strong assumptions on the redshift dependence of $\epsilon(z)$, one could exploit parameterizations of the redshift trend of such a function \citep{Lv:2016mmq}. Instead, we consider a simple binning of this function in two redshift bins, that is
\begin{equation}\label{eq:epsbin}
    \epsilon(z)= \begin{cases} \epsilon_0 &\text{if }z<z_*\,, \\ \epsilon_1 &\text{if }z\ge z_*\,, \end{cases}
\end{equation}
where $z_*$ is a transition redshift; we will comment on the choice of this redshift in the results section. 

In order to constrain these two parameterizations, we implement them in a new likelihood module interfaced with the publicly available MCMC sampler \texttt{Cobaya} \citep{Torrado:2020dgo}, able to reconstruct the posterior distribution of cosmological parameters, using SnIa and BAO data coming from current surveys or from simulated datasets. SnIa data are compared with the theoretical predictions given by \Cref{eq:epsz}, while with BAO data we compare combinations of the Hubble parameter $H(z)$ and of the standard angular diameter distance of \Cref{eq:angdist}.

We assume for this parameterized approach that the Universe expansion is well described by a flat $\Lambda$CDM model, with the late time evolution dominated by a cosmological constant with equation of state parameter $w(z)=-1$.
Given the flatness assumption, the angular diameter distance appearing in \Cref{eq:eps0} can be obtained in terms of the Hubble parameter $H(z)$ as
\begin{equation}\label{eq:angdist}
    d_{\rm A}(z) = \frac{c}{1+z}\int_0^z{\frac{\text{d}z'}{H(z')}}\, .
\end{equation}
 Therefore, we sample through \texttt{Cobaya} $\epsilon_0$ and $\epsilon_1$, parameterizing deviations from the standard DDR, alongside the total energy density of matter $\Omega_{\rm m,0}$, and the Hubble constant $H_0$, using flat priors on these parameters. The assumption of a flat Universe implies that the energy density given by the cosmological constant $\Lambda$ is $\Omega_{\rm \Lambda,0}=1-\Omega_{\rm m,0}$, neglecting the contribution of radiation energy density since we are analysing low-redshift data. Furthermore, we fix the baryon energy density to the mean obtained by Planck $\Omega_{\rm b,0} h^2=0.02225$ \citep{Aghanim:2018eyx}. Abandoning the assumption of $\Lambda$CDM and allowing for free parameters describing the equation of state of the new dark energy component would impact the constraints on DDR violation parameters, with the possibility of introducing degeneracies between the parameters determining $w(z)$ and $\epsilon(z)$; we leave however the investigation of this possibility for future work.

\subsection{Genetic algorithms}\label{sec:GA}

The GAs represent a class of machine learning methods that can be used for non-parametric reconstruction of data and are based on the notions of grammatical evolution, as expressed by the genetic operations of crossover and mutation. In particular, the GAs mimic the principle of evolution through the implementation of natural selection; a group of individuals evolves over time under the influence of the stochastic operators of mutation, namely a random change in an individual, and crossover, that is the combination of different individuals to form offspring. 

The probability that a member of the population will produce offspring, or in other terms its “reproductive success”, is assumed to be proportional to its fitness. The latter measures how accurately each individual of the population fits the data, here quantified through a $\chi^2$ statistic \citep[for more details on the GA and various applications to cosmology see][]{Bogdanos:2009ib,Akrami:2009hp,Nesseris:2010ep,Nesseris:2012tt,Nesseris:2013bia,Sapone:2014nna, Arjona:2019fwb,Arjona:2020kco,Arjona:2020doi}, which is obtained following the same likelihood computation used in \Cref{sec:parapp}. 

Qualitatively, the joint reconstruction of the SnIa and BAO data with the GA proceeds as follows. 
An initial population of functions is randomly chosen such that every member of the population contains initial guesses for both the luminosity distance $d_\textrm{L}(z)$ and the duality parameter $\eta(z)$.  
At this point we also impose some physical priors, such as that the luminosity distance at $z=0$ is zero, but we make no assumption on a DE model. Then, each member's fitness is calculated via a $\chi^2$ statistic, using as input the SnIa and BAO data and their individual covariances. Subsequently, the mutation and crossover operators are applied to the best-fitting functions in every generation, chosen via tournament selection---see \citet{Bogdanos:2009ib} for more details. This process is then iterated thousands of times, so as to ensure convergence, and with different random seeds, so as not to bias the results due to a specific choice of the random seed. 

After the GAs code has converged, the final output is a pair of two continuous and differentiable functions of redshift that describe the luminosity distance $d_\textrm{L}(z)$ and the duality parameter $\eta(z)$, respectively. At every step the angular diameter distance is calculated following \Cref{eq:eta1}, while the Hubble parameter $H(z)$ is calculated via differentiation of the latter assuming flatness. In the case of the current data we also numerically minimize the $\chi^2$ at every step over the combination $r_{\rm s}(z_{\rm d}) h$, with $r_{\rm s}(z_{\rm d})$ the comoving sound horizon at the drag epoch and $h=H_0/(100\ {\rm km\ s^{-1}\ Mpc^{-1}})$, in order to avoid making any model assumptions for the BAO physics at early times.

To estimate the errors on the reconstructed functions, we use an analytical approach developed by \citet{Nesseris:2012tt,Nesseris:2013bia}, where the errors are calculated via a path integral over the whole functional space that can be scanned by the GA. The GA path integral approach was extensively tested by \citet{Nesseris:2012tt} and found to be in excellent agreement with bootstrap Monte-Carlo error estimates.

In summary, using this approach we can reconstruct any cosmological function, for example the luminosity distance $d_{\rm L}(z)$ or the duality parameter $\eta(z)$ that we consider here, by applying the GA to any dataset of choice. No assumptions on the specific cosmological model or the behaviour of DE need to be made, hence our results are independent from specific DDR violation models. 
Since in our case the best-fit is very close to $\Lambda$CDM and the errors are much larger than the effects of any possible model-bias in the covariances of the data, we can safely assume for the time being that these effects have a rather minimal impact on the whole minimization process.

Finally, for the numerical implementation of the GA used in this paper we use the publicly available code \texttt{Genetic Algorithms}\footnote{\url{https://github.com/snesseris/Genetic-Algorithms}}. In addition to performing a large number of GA runs with different random seed numbers, we have also required that all reconstructed functions, as well as their derivatives, are continuous in the range of redshifts we consider, in order to avoid spurious reconstructions and overfitting.

\section{Analysis of currently available data}\label{sec:curr_data}

In order to constrain the deviation from the standard DDR, we need to analyse a set of data providing information on the luminosity and angular diameter distances. We focus therefore on currently available observations of SnIa and BAO.

The BAO data will provide information on the angular diameter distance $d_{\rm A}(z)$ and the Hubble parameter $H(z)$. We use here measurements of the ratio $d_z$, defined as 
\be\label{eq:dz}
d_z\equiv \frac{r_{\rm s}(z_{\rm d})}{D_V(z)},
\ee
where $D_V$ is the volume averaged distance 
\be 
D_V(z)=\left[(1+z)^2 d_{\rm A}^2(z) \frac{c z}{H(z)}\right]^{1/3}\, ,
\ee 
and $r_{\rm s}(z_{\rm d})$ is the comoving sound horizon at the drag epoch 
\be
r_{\rm s}(z_{\rm d})=\frac{1}{H_0}\int_{z_{\rm d}}^\infty \frac{c_{\rm s}(z)}{H(z)/H_0} \,\text{d}z\, ,
\label{eq:sound-horizon-drag}
\ee
with $c_{\rm s}(z)$ the sound speed and $z_{\rm d}$ the redshift at the drag epoch \citep[see Eq. 4 of][]{Eisenstein:1997ik}. In the \lcdm model, \Cref{eq:sound-horizon-drag} can be approximated as \citep[see Eq. 26 of][]{Eisenstein:1997ik}
\be 
r_{\rm s}(z_{\rm d})\simeq \frac{44.5\log\left(\frac {9.83} {\Omega_{\rm m, 0}h^2}\right)}{\sqrt {1+10(\Omega_{\rm b, 0}h^2)^{3/4}}} ~\textrm{Mpc}\, .
\ee
Throughout this paper we will assume that this approximation holds in all our parameterized analyses. Moreover, as the data combination considered here cannot constrain $\Omega_{\rm b, 0}h^2$, we assume the value $\Omega_{\rm b, 0} h^2=0.02225$ from Planck 2018 \citep{Aghanim:2018eyx}.
We notice that the constraints one can obtain through analysis of the BAO can depend significantly on this assumption, with different choices available on how to obtain prior information on $r_{\rm s}(z_{\rm d})$ \citep[see for example][for a detailed discussion on the role of $r_{\rm s}(z_{\rm d})$ assumptions in BAO analysis]{Cuesta:2014asa}. For instance, a change of $1\%$ in the value of $\Omega_{\rm b,0}h^2$ leads to a change of about $2\%$ on the distance ratio $d_z$. The observational constraints, on the other hand, on the quantity given by \Cref{eq:dz}, as well as on the Hubble distance $D_{\rm H}(z)=c/H(z)$, that we consider here are provided by the surveys 6dFGS \citep{Beutler:2011hx}, SDDS \citep{Anderson:2013zyy}, BOSS CMASS \citep{Xu:2012hg}, WiggleZ \citep{Blake:2012pj}, MGS \citep{Ross:2014qpa}, BOSS DR12 \citep{Gil-Marin:2015nqa}, DES \citep{Abbott:2017wcz}, Ly-$\alpha$ observations from \citet{Blomqvist:2019rah}, SDSS DR14 LRG \citep{Bautista:2017wwp} and quasars observations from \citet{Ata:2017dya}. In the rest of the paper, we will refer to the combination of these datasets as BAO, for simplicity. We refer the reader to \Cref{app:BAOchi} for further details on how these datasets are combined together and a description of their likelihood.

We also note that the BAO data may have some model dependence, as a fiducial cosmology is required in order to convert the measured angular scales to distances, while some more uncertainty may also be introduced by the non-linear effects which damp and modify the position of the BAO in the galaxy power spectrum. Both of these issues of course imply that systematic errors of a few percent may be introduced in the inferred cosmological parameters   \citep[see][]{Angulo:2007fw}. While it is possible to standardise the BAO distance measurements, many of these techniques are based on the particular modelling of non-linear scales, something which is quite complicated for theories beyond the $\Lambda$CDM model. Not including the non-linear modelling, may thus lead to reduced constraining power \citep[see for example][]{Anselmi:2017cuq}.

On the other hand, the SnIa data provide information on the luminosity distance $d_{\rm L}(z)$, as the measured observable is the apparent magnitude $m(z)$ which can be expressed as
\begin{equation}
    m(z) = M_0+5\log_{10}{\left(\frac{d_{\rm L}(z)}{\rm Mpc}\right)}+25\, ,
\end{equation}
where $M_0$ is the intrinsic magnitude of the considered supernova. Such a quantity is completely degenerate with the Hubble constant $H_0$, thus, if no external information is provided, SnIa data are not able to constrain these two quantities. Here, we analyse the SnIa data using the likelihood expression from Appendix C in \citet{Conley:2011ku}, which already takes into account the marginalization of $M_0$ and $H_0$ from the SnIa analysis. The dataset we consider for the SnIa is the updated Pantheon compilation of 1048 points from \citet{Scolnic:2017caz}. 

\subsection{Parameterized results}

\begin{figure}
    \centering
    \includegraphics[width=0.9\columnwidth]{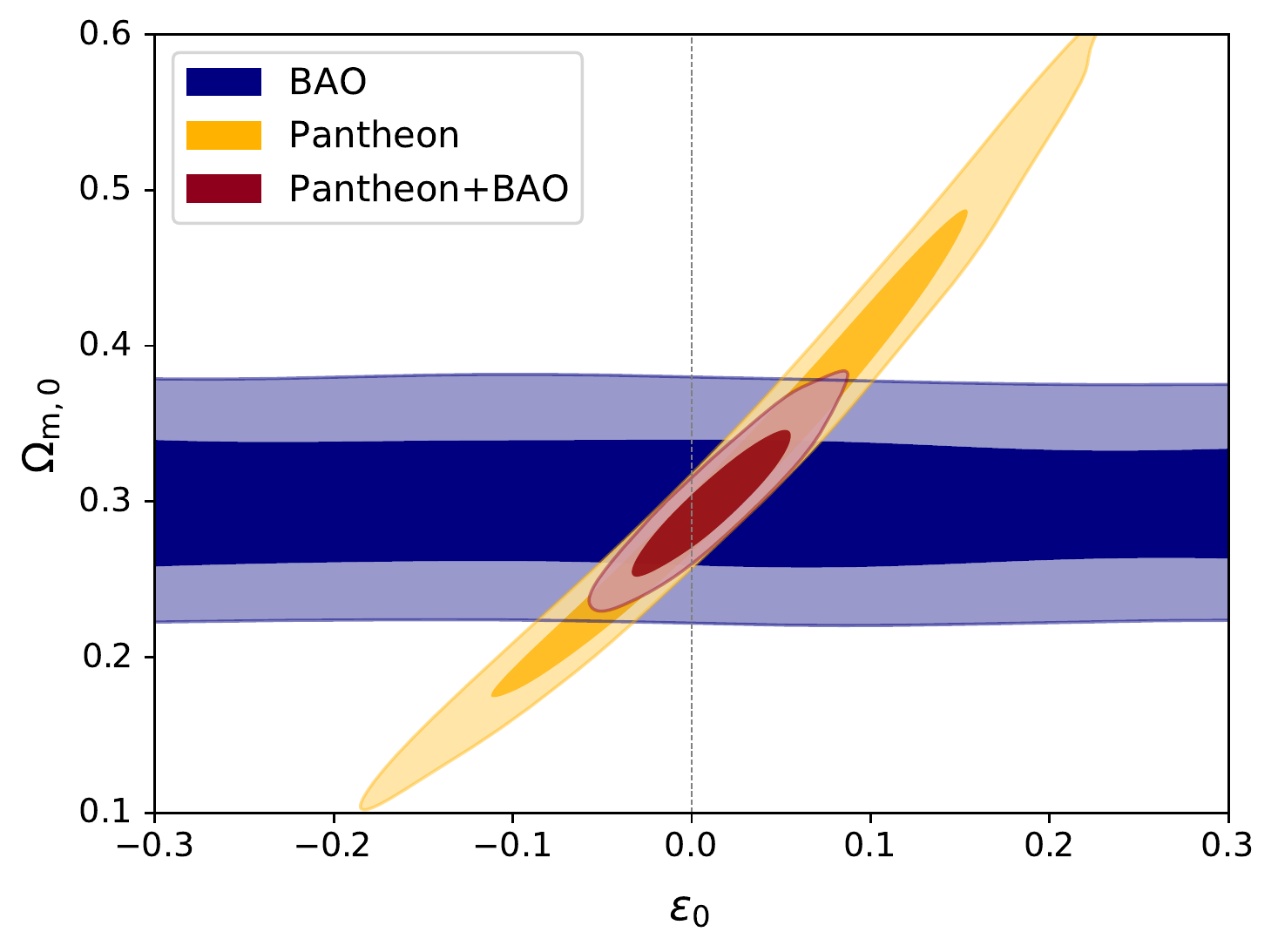} \\
    \includegraphics[width=0.9\columnwidth]{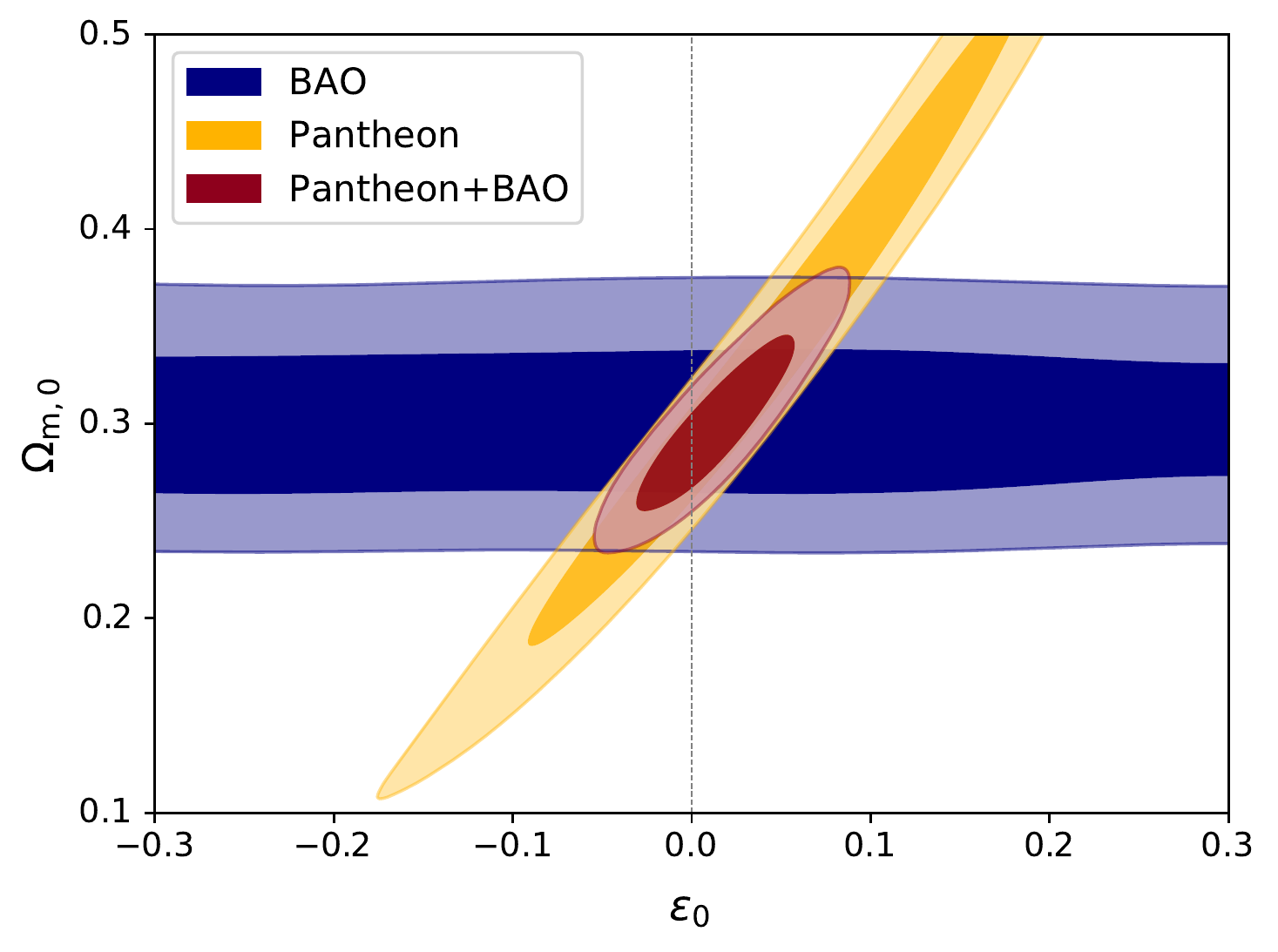} \\
    \includegraphics[width=0.9\columnwidth]{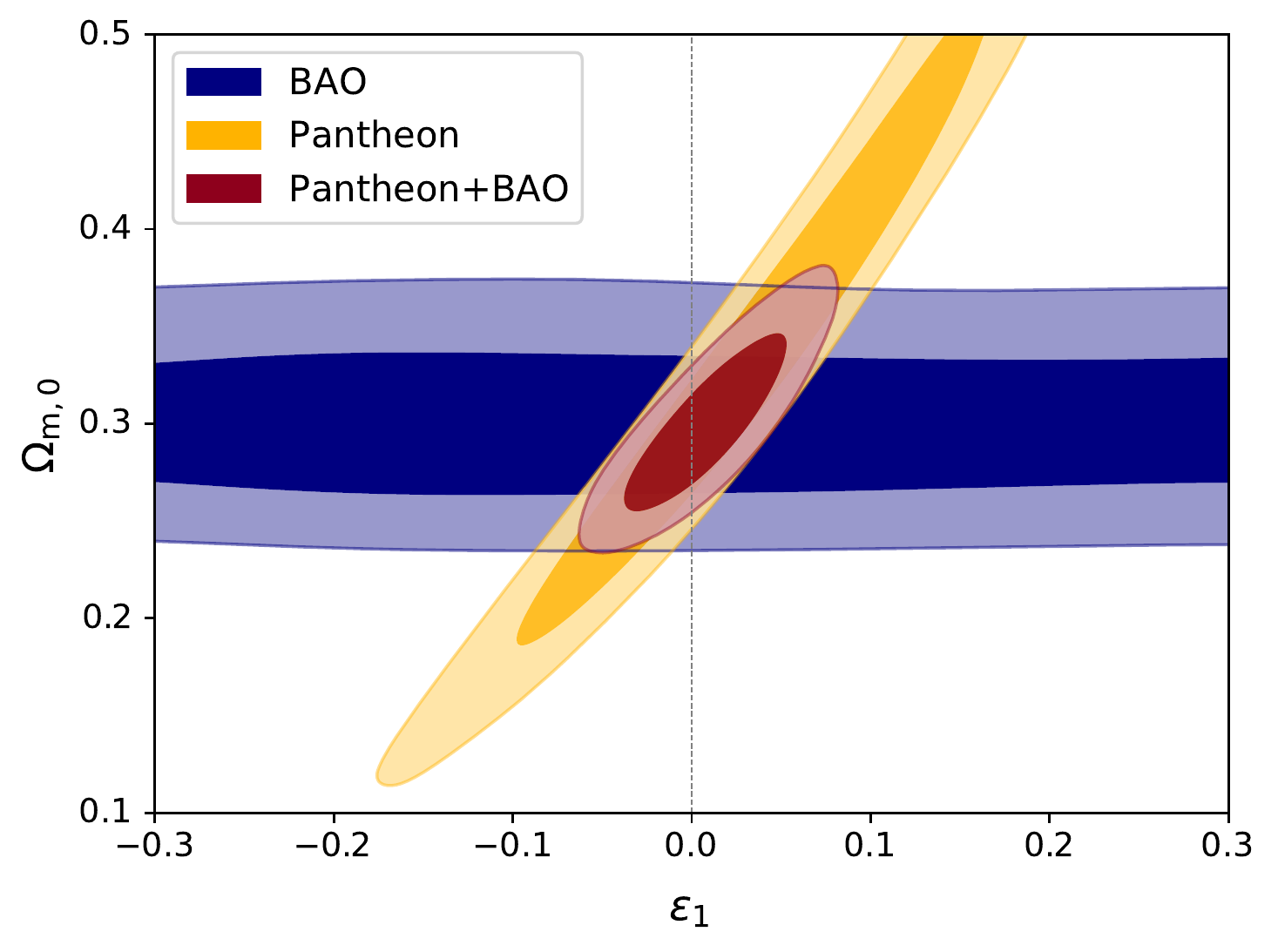} \\
    \caption{2D contours on $\Omega_{\rm m,0}$, $\epsilon_0$ and $\epsilon_1$, using currently available data for BAO (blue), SnIa (yellow) and the combination of the two (red). These results refer to the constant (top panel) and binned (central and bottom panels) $\epsilon(z)$ cases.}
    \label{fig:current_res}
\end{figure}

Using the surveys described above we can quantify the current constraining power on the DDR, both in the constant and binned cases of the $\epsilon$ parameterization. In \Cref{tab:current_res} we report both of these results, whereas in \Cref{fig:current_res} we show the constraints obtained from the considered observables, both separately and in combination. The contours shown here clarify how SnIa data are able to constrain $\epsilon(z)$, although the degeneracy between $\Omega_{\rm m,0}$ and the DDR parameters significantly limits the constraining power. The BAO data on the other hand, are not sensitive to $\epsilon(z)$, but are able to obtain tight constraints on the allowed matter density. Therefore, when the two datasets are combined the degeneracy between $\Omega_{\rm m,0}$ and $\epsilon_0,\ \epsilon_1$ is broken.

One may notice that the mean of the DDR parameters is positive, both in the constant and binned cases, and that the inclusion of BAO data shifts the constraints towards the standard cosmology limit $\epsilon(z)=0$. Using alternative BAO combinations, producing different constraints on $\Omega_{\rm m,0}$, can therefore lead to different results on $\epsilon(z)$; this is crucial if one wants to connect the constraints to viable theoretical models producing the inferred violation of DDR, as the mechanisms leading to a positive or negative $\epsilon(z)$ can be significantly different, as discussed in \Cref{sec:theory}.

Considering the combined Pantheon+BAO constraints, in the constant $\epsilon(z)$ case, the posterior distribution peaks at $\epsilon_0\ne0$; however, within the $1\sigma$ limit, the result is compatible with zero. The binned $\epsilon(z)$ case shows a similar behaviour: Both the first and second bin parameters, that is $\epsilon_0$ and $\epsilon_1$, are compatible with zero at 
$1\sigma$. Furthermore, the errors on these two parameters are very similar, showing how current data provide similar constraining power in the two redshift bins considered here. This is due to the fact that the transition redshift $z_*$, fixed here to $z_*=0.9$, lies roughly midway through the redshift range of the SnIa data, which are those sensitive to DDR parameters. It is important to stress here that throughout this analysis, we keep the transition redshift $z_*$ fixed. Moreover, current data do not provide any hint for $\epsilon_0\ne\epsilon_1$, as the constraints of both parameters are compatible with each other and therefore consistent with a constant $\epsilon(z)$. We analysed the data also allowing for a free $z_*$; however, as the data are compatible with a constant $\epsilon(z)$, no clear peak of the posterior distribution is present, with the two extreme cases $z_*\approx0$ and $z_*\gtrsim2$ providing the same result. Therefore, the posterior distribution is extremely difficult to sample with the MCMC algorithm used here and we decide to present here only the analysis where $z_*$ is fixed.

 In particular, using both the current SnIa and BAO data, we find $\epsilon_0=0.013\pm0.029$, while, when the binned approach is used, we find $\epsilon_1=0.009\pm 0.030$ and $\epsilon_0=0.015
^{+0.027}_{-0.031}$.
The former may be compared to the analysis by \citet{Avgoustidis2010}, where $\epsilon_0$ was found to be $\epsilon_0=-0.04^{+0.08}_{-0.07}$ (all of these being at the $68\%$ confidence level). 

\begin{figure}
    \centering
    \vspace{-0.5cm}
    \includegraphics[width=0.9\columnwidth]{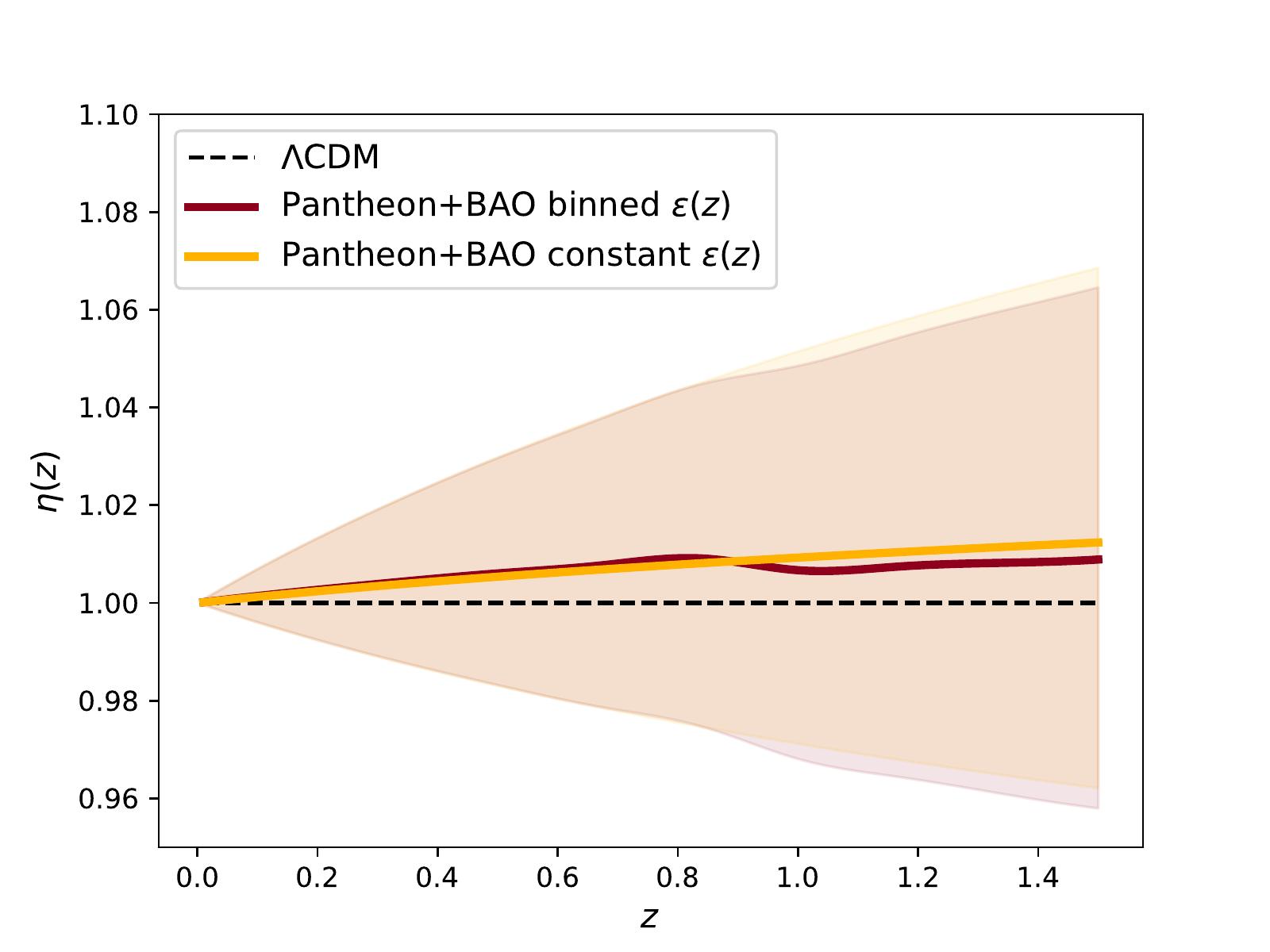}

    \caption{Reconstruction with current SnIa and BAO data of the $\eta(z)$ function at different redshifts as derived parameters using \Cref{eq:eta1}. The mean function is shown as a solid line, while the shaded area represents the $68\%$ confidence region. The red colour shows the result in the binned $\epsilon(z)$ case, while yellow refers to the constant case}
    \label{fig:current_eta}
\end{figure}

Finally, we show in \Cref{fig:current_eta} the reconstructed trend of $\eta(z)$, whose values at different redshifts are obtained as a derived parameter using \Cref{eq:eta1}. As can be seen, the reconstruction is in agreement with \lcdm within the errors. 

\begin{table}[!htbp]
\begin{center}
\caption{Mean values and marginalized $68\%$ confidence level errors obtained from currently available data on the cosmological parameters $\Omega_{\rm m,0}$ and $H_0$ (in units of km s$^{-1}$ Mpc$^{-1}$) and on the DDR parameters $\epsilon_0$ and $\epsilon_1$ (if present).}\label{tab:current_res}
\begin{tabular}{lc|c|c} 
\hline
 \multicolumn{2}{c}{}& constant $\epsilon(z)$ & binned $\epsilon(z)$ \\
\hline
parameter & probe & & \\
\hline
             & BAO & $66.6^{+1.3}_{-1.4}$ & $66.7\pm 1.3$\\
$H_0$        & SnIa & unconstrained & unconstrained\\
             & SnIa+BAO & $66.6\pm 1.3$ & $66.6\pm 1.3$\\
\hline
             & BAO & $0.300^{+0.027}_{-0.036}$ & $0.302^{+0.027}_{-0.035}$ \\
$\Omega_{\rm m,0}$   & SnIa & $0.329^{+0.094}_{-0.12}$ & $0.357^{+0.090}_{-0.14}$ \\
             & SnIa+BAO & $0.301^{+0.028}_{-0.034}$ & $0.301^{+0.026}_{-0.033}$\\
\hline
             & BAO & unconstrained & unconstrained\\
$\epsilon_0$ & SnIa & $0.030\pm 0.088$ & $0.056^{+0.087}_{-0.10}$\\
             & SnIa+BAO & $0.013\pm 0.029$ & $0.015^{+0.027}_{-0.031}$\\
\hline
             & BAO & --- & unconstrained\\
$\epsilon_1$ & SnIa & --- & $0.046\pm 0.089$\\
             & SnIa+BAO & --- & $0.009\pm 0.030$\\
\hline
\hline 
\end{tabular}
\end{center}
\end{table}

\subsection{GA sresults}
In order to obtain constraints on the violation of the DDR without assuming any specific trend in redshift for $\epsilon(z)$, we employ here the GAs approach described in \Cref{sec:GA}, applied simultaneously to the currently available SnIa and BAO data. We find that a joint fit with the GAs to both data sets gives a competitive fit with respect to the \lcdm model. In particular, after applying the GAs we find a best-fit of $\chi^2_{\rm min,GA}=1041.510$ for $r_{\rm s}(z_{\rm d})=100.360~\textrm{Mpc}/h$. Concerning the \lcdm model, 
we obtain a minimum value of $\chi^2_{\rm min,\Lambda \textrm{CDM}}=1045.696$, for $1048+12=1060$ data points (1048 from the SnIa and 12 from the BAO), for the best-fit matter density parameter $\Omega_{\rm m,0}=0.297\pm0.018$ and $H_0=66.7\pm 1.0~ \textrm{km/s/Mpc}$. Overall, the GA provides a better fit to the data, with $\Delta \chi^2 = 4.187$, compared to the \lcdm model. 

As mentioned before, the output of the GAs is an analytical function, but  in most cases the exact expression is both cumbersome and not informative. Though, in this case we were able to find a compact expression for the GA reconstruction of the $\eta(z)$ parameter, given by
\be
\eta(z)=(1+z)^{0.0294 - 0.0002\;z^4}.\label{eq:etaGA}
\ee
As can be seen, the value of $\epsilon(z)$ derived from \Cref{eq:etaGA} is compatible with the one derived in the parameterized approach. Moreover, the value predicted from the GAs has a redshift dependence $\mathcal{O}(z^4)$, which is only important at high redshifts as the coefficient is sufficiently small, albeit negative. This is also in agreement with the fact that the parameterized approach finds a value for $\epsilon$ that is smaller in the second bin with respect to the first. Finally, the GAs reconstructions for the luminosity distance or the Hubble parameter are unfortunately far too unwieldy, so we refrain from reporting them here.

\begin{figure*}[!t]
	\centering
	\includegraphics[width=0.49\textwidth]{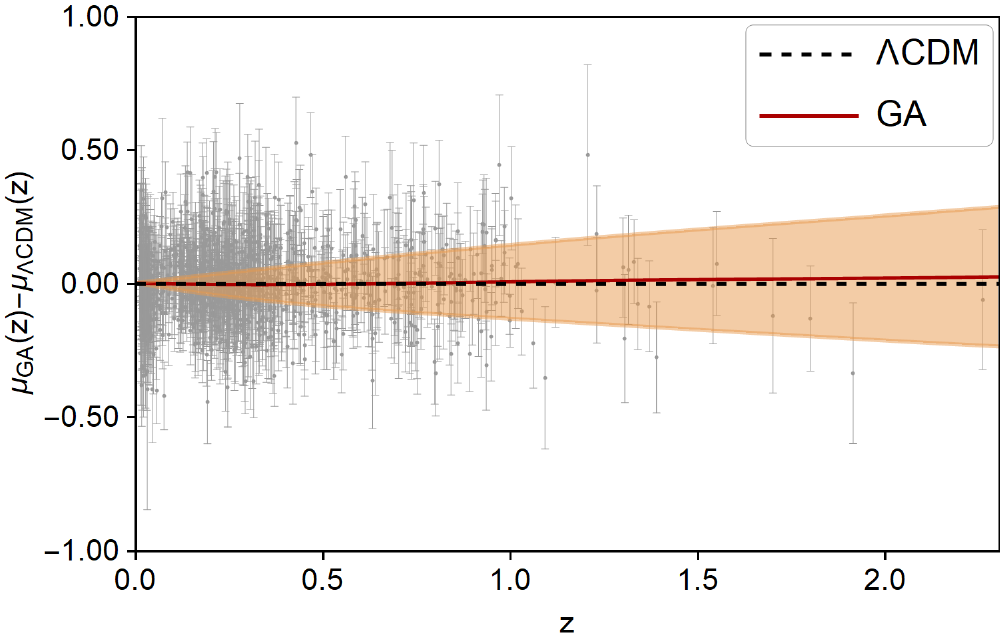}\hspace{.3cm}
	\includegraphics[width=0.482\textwidth]{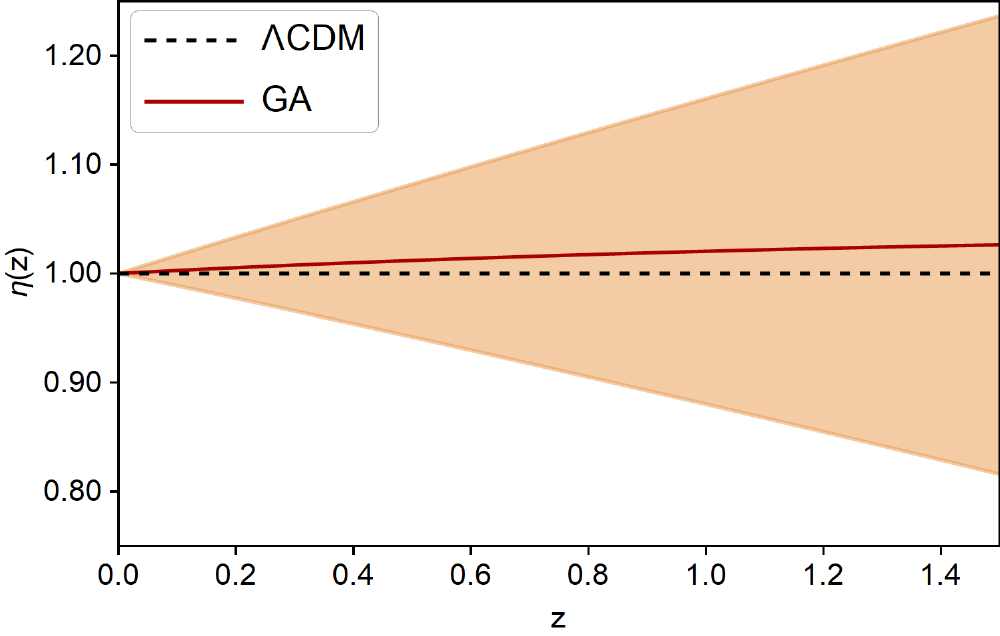}
	\caption{Left: Distance modulus based on the Pantheon SnIa set, rescaled by the best-fit $\Lambda$CDM model ($\Omega_{\rm m,0}=0.297\pm0.018$). The dashed line at zero corresponds to $\Lambda$CDM, the red line is the GA fit, and the shaded region corresponds to the $1\sigma$ GA errors. Right: Duality relation $\eta(z)$ for the GA reconstruction of $d_{\rm A}(z)$ from the BAO data and the luminosity distance $d_{\rm L}(z)$ based on the Pantheon SnIa set. The dashed line at unity is $\Lambda$CDM,  the red line is the GA fit, and the shaded region is the $1\sigma$ GA errors.  \label{fig:realplotsGA}}
\end{figure*}

Having performed the fit to the data, we now show in the left panel of \Cref{fig:realplotsGA} the reconstruction of the distance modulus $\mu(z)=m(z)-M_0$, rescaled by the $\Lambda$CDM best-fit with $\Omega_{\rm m,0}=0.297\pm0.018$. By definition, the dashed line at zero corresponds to the best-fit $\Lambda$CDM, the red line is the GAs fit and the shaded region corresponds to the $1\sigma$ GA errors. The constraints on the distance modulus are tighter at low redshifts $z\in[0,0.5]$ where we have the bulk of the BAO and SnIa data points, but also due to the fact that the distance modulus $\mu(z)$ is a function of the luminosity distance $d_\textrm{L}(z)$ and as a result it naturally converges to a fixed value at $z=0$.

In the right panel \Cref{fig:realplotsGA} we show the duality parameter $\eta(z)$ of \Cref{eq:eta1}, obtained through the GAs reconstruction of $d_{\rm A}(z)$ from the BAO data and the luminosity distance $d_{\rm L}(z)$ based on the Pantheon SnIa set. The dashed line at unity is  $\Lambda$CDM, the red line is the GAs fit and the shaded region corresponds to the $1\sigma$ GAs errors. As can be seen, the reconstruction of  $\eta(z)$ is compatible with unity at the $1\sigma$ level, hence the GAs do not detect any statistically significant deviations from the \lcdm model with the currently available data.

\section{SnIa and BAO mock data}\label{sec:fore_data}

Upcoming surveys have the potential to improve the constraining power on deviations from the standard DDR, thanks to improved data both for SnIa and LSS observations, where the latter provide information on the angular diameter distance and the Hubble parameter. We are interested therefore in forecasting how future surveys will constrain the DDR and in order to do so we create simulated datasets for both SnIa and BAO measurements.

For these simulations, we use the fiducial cosmology shown in \Cref{tab:fiducials}, that is the same used in \citetalias{IST:paper1}, where we assume no violation of the DDR. Using these values we create our fiducial luminosity and angular diameter distances, as well as the redshift evolution of the Hubble parameter. Once the fiducial cosmological quantities are computed, we create our mock data following the specification of forthcoming surveys. 

It is important to mention that given the high precision expected from future surveys, it will be even more important to ensure their accuracy. Several analyses are being performed to understand the observational systematic uncertainties that will affect future measurements. In \citet{Paykari2020}, for instance, a detailed analysis on the observational systematic effects related to the \Euclid VIS instrument is performed, including charge transfer ineffiency and modelling of the point spread function. Given that the specifications for the future instruments and the modelling of their systematic effects might still evolve during their completion, in this work we assume that the observational systematic effects will be under control when the data arrive. Nevertheless, we do include astrophysical systematic effects, like galaxy bias, as described in the following subsections.

\subsection{SnIa surveys}
Here we consider two different surveys. On the one hand, we simulate future observations based on the specifications of the LSST, which we assume will observe a number of SnIa $N_{\rm SnIa}=8800$ in the redshift range $z\in[0.1,1.0]$. We then extend the redshift range of our SnIa dataset by including simulated observations for the proposed \Euclid DESIRE survey \citep{Laureijs:2011gra,Astier:2014swa}, thus including $1700$ additional data points in the range $z\in[0.7,1.6]$. For both surveys, we assume the redshift distributions shown in \citet{Astier:2014swa} and we further assume that the two are not correlated\footnote{We stress that the DESIRE survey is not a guaranteed output of \Euclid. Here we include this in the analysis as a possible survey extending the redshift range of LSST. Such a survey will be crucial for performing the GA reconstruction at higher redshifts.}. For each event, we simulate an observational error $\sigma_{{\rm tot},i}$ given by
\begin{equation}
    \sigma_{\textrm{tot},i}^2=\delta \mu^2_i+\sigma^2_{\textrm{flux}}+\sigma^2_{\textrm{scat}}+\sigma^2_{\textrm{intr}}\,,
\end{equation}
where the flux, scatter, and intrinsic contributions are the same for each event ($\sigma_{\textrm{flux}} = 0.01$, $\sigma_{\textrm{scat}} = 0.025$, and $\sigma_{\textrm{intr}} = 0.12$, respectively) and we add an error contribution on the distance modulus $\mu=m-M$, which evolves linearly in redshift
\begin{equation}
    \delta\mu = e_M~z,
\end{equation}
where $e_M$ is drawn from a Gaussian distribution with vanishing mean and $\sigma(e_M)=0.01$ \citep[see][]{Gong:2009yk,Astier:2014swa}.

We note that while the effects of lensing by foreground structures are already included in the Pantheon data by incorporating an error  $\sigma_\textrm{lens}$  \citep[see][]{Scolnic:2017caz}, here we have not included this error in our mocks as it has a very weak redshift dependence and is subdominant with respect to the intrinsic distance scatter of every point of $\sim0.12$mag. Thus, we do not expect it to affect our results.

\subsection{LSS surveys}
One of the main objectives of this paper is to forecast the constraints achievable on DDR with \Euclid. As such, we simulate BAO data from this survey using the Fisher matrix technique, following the same strategy used in \citetalias{IST:paper1} for the spectroscopic survey. 

Since in this work we are interested in using precise measurements of the Hubble parameter and the angular diameter distance to test the DDR, we will focus on the spectroscopic \Euclid survey. Through this,  \Euclid will be capable of exploring the galaxy power spectrum in a range of redshifts $z \in [0.95,1.75]$. As described in \citetalias{IST:paper1}, the main targets are ${\rm H}_{\alpha}$ emitters and the survey is able to measure up to $30$ million spectroscopic redshifts with an error of  $\sigma_z = 0.001(1 + z)$ \citep{Pozzetti:2016cch}. The main observable is the galaxy power spectrum which contains information about the galaxy bias, the anisotropies due to redshift space distortions, the residual shot noise, the redshift uncertainty and the distortion due to the Alcock-Paczynski effect. Furthermore, the matter power spectrum has been modulated with non-linear effects which distort the shape of the power spectrum \citep{Wang:2012bx}.

With respect to \citetalias{IST:paper1}, in this work we use a different binning scheme. Instead of four redshift bins we divide the observed redshift range in nine equally spaced bins of width $\Delta z = 0.1$.
The galaxy number density $n(z)$, in units of Mpc$^{-3}$ and the galaxy bias $b(z)$ have been obtained rebinning those of \citetalias{IST:paper1}, finding: 
\begin{eqnarray}
n(z)&=& \{2.04, 2.08, 1.78, 1.58, 1.39, 1.15, 0.97, 0.7, 0.6\}\times 10^{-4}\,\nonumber \\
b(z)&=& \{1.42, 1.5, 1.57, 1.64, 1.71, 1.78, 1.84, 1.90, 1.96\}.\nonumber
\end{eqnarray}
The different binning choice allows obtaining more data points from this survey, which improves the machine learning analysis we perform through GA. Nevertheless, we have compared the final bounds obtained on cosmological parameters with this choice against those of \citetalias{IST:paper1}, finding no significant effect. 

Using these specifications, we follow the procedure described in \citetalias{IST:paper1} to obtain the Fisher matrix for the full set of cosmological parameters, namely: four shape parameters $\{\omega_{\rm m}=\Omega_{\rm m,0}h^2$, $h$, $\omega_{\rm b}=\Omega_{\rm b,0}h^2$, $n_{s}\}$, two non-linear parameters $\{\sigma_{\rm p},\,\sigma_{\rm v}\}$ and five redshift dependent parameters $\{\ln d_{\rm A},\,\ln H,\,\ln f\sigma_8,\,\ln b\sigma_8,\,P_{\rm s}\}$ evaluated in each redshift bin. Using such an approach, we obtain the expected errors from this survey on the angular diameter distance $d_{\rm A}(z)$ and the Hubble parameter $H(z)$ in each of the nine redshift bins, while marginalizing over all the other free parameters. The results of the Fisher matrix procedure are in principle dependent on the chosen fiducial cosmology, but we assume here that this dependence is negligible.

In \Cref{sec:curr_data} we have shown how for our parameterized approach we need to break the degeneracy between the DDR parameters and $\Omega_{\rm m,0}$, and the BAO measurements from \Euclid will be able to measure this parameter. However, we are able to use our GA reconstruction approach only in the redshift range where both SnIa and BAO data are available. Using only \Euclid alongside LSST and DESIRE would therefore limit the validity of such an approach to only the redshift range $z \in [0.95,1.6]$. In order to be able to reconstruct the DDR functions at all redshift for which we have SnIa data available, we complement the redshift range of \Euclid by exploiting the extended redshift range of the DESI survey, which started operations at the end of 2019 and will obtain optical spectra for tens of millions of galaxies and quasars up to redshift $z\sim 4$. 

Such spectra will enable BAO and redshift-space distortion cosmological analyses. We use here the official DESI forecasts on future constraints for both $H(z)$ and $d_{\rm A}(z)$ \citep{DESI2016}. These have been obtained with a Fisher matrix formalism, following \citet{2014JCAP...05..023F}, which includes the 'broadband' galaxy power, meaning measurements of the power spectrum as a function of redshift, wavenumber, and angle with respect to the line of sight. As for the \Euclid approach described above, this encodes all the available information from the two-point clustering and not just the position of the BAO peak. In more detail, we consider the DESI baseline survey, which consists of a coverage of 14\,000\,deg$^2$ and the four different types of DESI targets: bright galaxies (BGs), luminous red galaxies (LRGs), emission line galaxies (ELGs), and quasars. The DESI forecast measurements will cover the redshift range $z\in [0.05,3.55]$, but their precision will also depend on the target population. The BGs will cover the redshift range $z\in [0.05,0.45]$ in five equispaced redshift bins, the LRGs and ELGs will focus on $z\in [0.65,1.85]$ with $13$ equispaced redshift bins, while the Ly-$\alpha$ forest quasar survey will cover $z\in [1.96,3.55]$ with $11$ equispaced redshift bins. We further assume these measurements to be uncorrelated.

In the following, when using the combination of BAO data from \Euclid and DESI, as we do not consider correlations between these surveys, we will only include DESI observations that do not overlap in redshift with the \Euclid measurements. Moreover, since we only have SnIa data from LSST+DESIRE up to $z=1.6$, we will only include in the analysis the full BGs survey and the LRGs and ELGs up to $z=0.9$, thus including no information from observations of the Ly-$\alpha$ forest.

\begin{table}
\begin{center}
\caption{Parameter values for the fiducial model we used for the mock. The values used follow the fiducial of \citetalias{IST:paper1}. $H_0$ is shown in units of km s$^{-1}$ Mpc$^{-1}$. \label{tab:fiducials}}
\begin{tabular}{cccccccc}
\hline
\hline
$M_0$ & $\Omega_{\rm m,0}$ & $\Omega_{\rm b,0}h^2$ & $H_0$ & $w_0$ & $w_a$ & $\epsilon_0$ & $\epsilon_1$ \\
 \hline
$-19.3$ & $0.32$ & $0.02225$ & $67$ & $-1$ & $0$ &$0$ &$0$ \\
\hline
\hline
\end{tabular}\\
\end{center}
\end{table}

\section{Forecast results}\label{sec:fore_res}

\begin{figure}
    \centering
    \includegraphics[width=0.9\columnwidth]{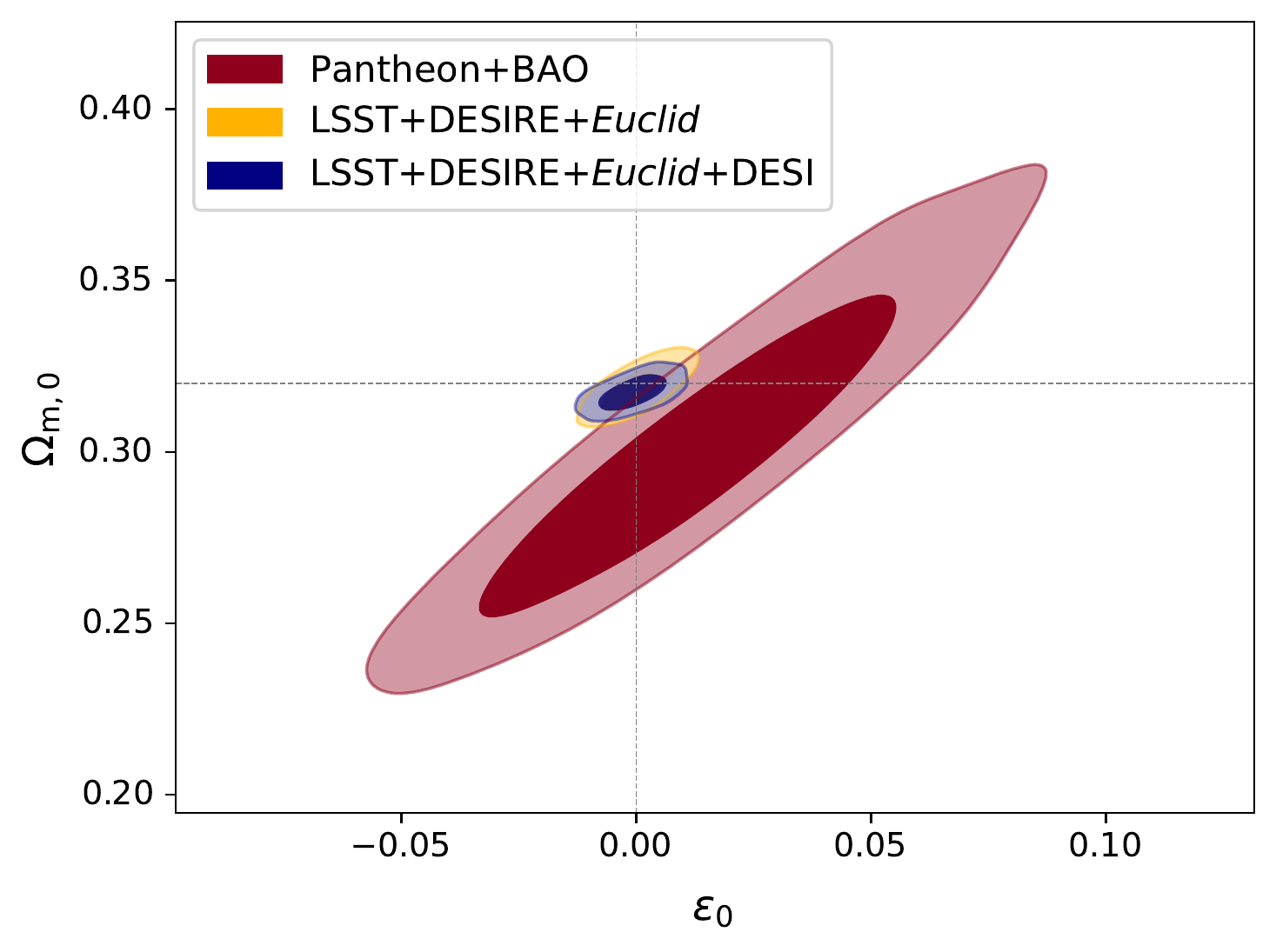} 
    \includegraphics[width=0.9\columnwidth]{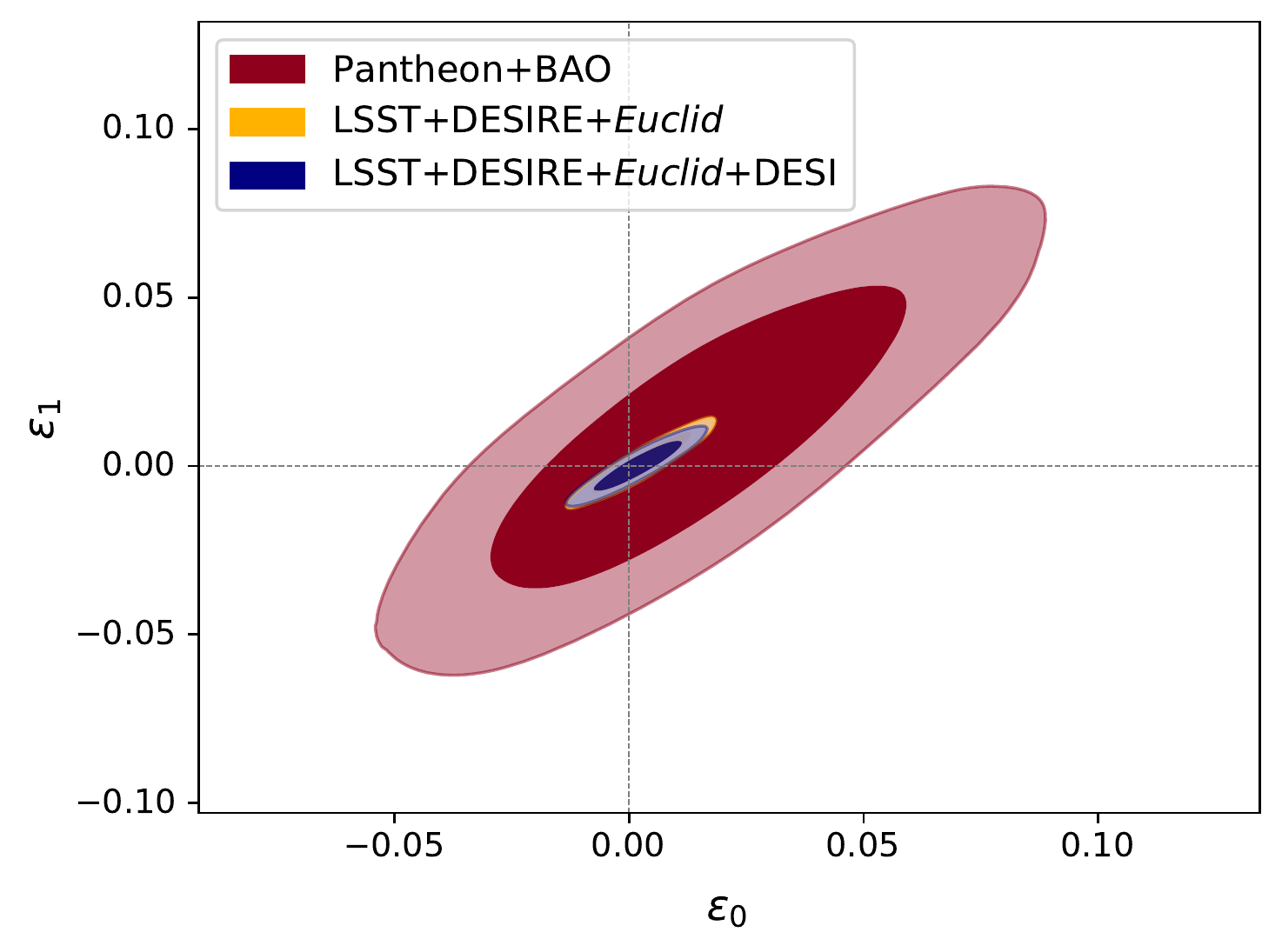}
    \caption{2D contours on $\Omega_{\rm m,0}$, $\epsilon_0$ and $\epsilon_1$, using the combination of BAO and SnIa dataset given by currently available data (red contours), LSST supernovae and \Euclid BAO data (yellow contours), and the combination of LSST supernovae with BAO forecasts coming from the combination of \Euclid and DESI (blue contours). These results refer to the constant (top panel) and binned (bottom panel) $\epsilon(z)$ cases. The dashed lines identify the limit $\epsilon(z)=0$.}
    \label{fig:forecasted_res}
\end{figure}

Following the approach described in \Cref{sec:analysis}, we constrain the cosmological and DDR parameters using our mock data for SnIa and BAO. In \Cref{tab:forecasted_res} we show the mean values and errors for the free parameters of the analysis, when using the data from LSST+DESIRE for SnIa and from {\it Euclid}+DESI for BAO, for both the constant and binned $\epsilon(z)$. In \Cref{fig:forecasted_res} instead we compare the results of the combination of these surveys with those obtained using current data and to what can be achieved using only the BAO survey from \Euclid.  This is to be compared with Fig. 54 in \citet{ReviewDoc}, whose forecast is for \Euclid (with the specifications foreseen at the time) plus a Stage IV (SNAP-like) SnIa mission: the achieved constraints are compatible with what we find here. In summary, $\epsilon_0$ and $\epsilon_1$ are now constrained with an error smaller than $10^{-2}$, improving the sensitivity of current constraints by about a factor of $\approx6$. We notice that here we are not considering one of the two primary probes of \Euclid, that is cosmic shear. Adding the information brought by such a probe would further constrain the value of $\Omega_{\rm m,0}$, thus resulting in even tighter bounds on the DDR parameters $\epsilon_0$ and $\epsilon_1$.

As can be seen in  \Cref{fig:forecasted_res} the addition of DESI to the combination of LSST+\Euclid does not improve the constraints significantly. This could be somewhat surprising, as one would expect that, given the complementarity in redshift range between the two surveys, the combination of the two would provide improvements in the constraints. Our results show instead that the constraining power on $\Omega_{\rm m,0}$ from \Euclid alone is the one dominating the constraints and therefore driving the breaking of the degeneracy between $\Omega_{\rm m,0}$ and $\epsilon(z)$ parameters. While the $z=0.9$ cut we perform to avoid an overlap of the two BAO surveys is not \textit{per se} behind this effect, we do expect that the use of the full DESI data (specifically the high-redshift galaxy and the Ly-$\alpha$ data) would lead to a stronger improvement in the constraints. However, in order to investigate this further, correlations between the two mock datasets should be properly taken into account, an analysis that is outside the scope of this paper.

\begin{figure}
    \centering

    \includegraphics[width=0.9\columnwidth]{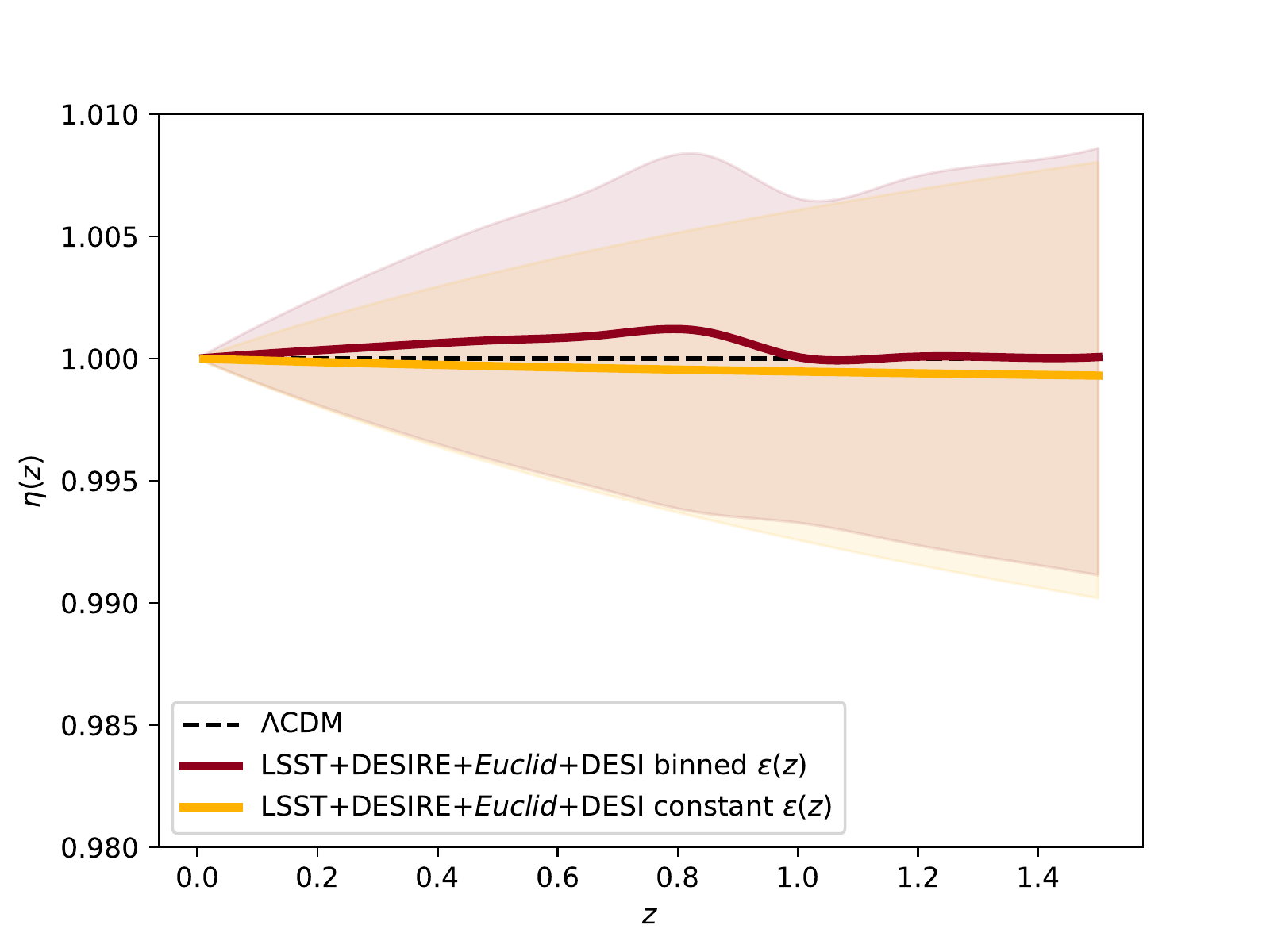}

    \caption{Reconstruction with future SnIa and BAO data of the $\eta(z)$ function at different redshifts, as derived parameters using \Cref{eq:eta1}. The mean function is shown as a solid line, while the shaded area represents the $68\%$ confidence region. The red colour shows the result in the binned $\epsilon(z)$ case, while yellow refers to the constant case.}
    \label{fig:forecast_eta}
\end{figure}

Furthermore, in \Cref{fig:forecast_eta} we show $\eta(z)$ obtained using \Cref{eq:eta1} with the parameterized approach constraints; we find that these are significantly improved with respect to current results, showing the high constraining power that can be reached using upcoming surveys.

\begin{table}[!tbp]
\begin{center}
\caption{Mean values and marginalized $68\%$ confidence level errors obtained from mock LSST (SnIa), \Euclid (SnIa and BAO) and DESI (BAO) data on the cosmological parameters $\Omega_{\rm m,0}$ and $H_0$ (in units of km s$^{-1}$ Mpc$^{-1}$) and on the DDR parameters $\epsilon_0$ and $\epsilon_1$ (if present).\label{tab:forecasted_res}}
\begin{tabular}{lc|c|c} 
\hline
 \multicolumn{2}{c}{}& const. $\epsilon(z)$ & binned $\epsilon(z)$ \\
\hline
param. & probe & & \\
\hline
             & BAO & $67.13\pm 0.25$ & $67.14\pm 0.25$\\
$H_0$        & SnIa & unconstrained & unconstrained\\
             & SnIa+BAO & $67.14\pm 0.26$ & $67.15\pm 0.26$\\
\hline
             & BAO & $0.3175\pm 0.0034$ & $0.3174\pm 0.0034$\\
$\Omega_{\rm m,0}$   & SnIa & $0.259\pm 0.077$ & $0.281\pm 0.096$\\
             & SnIa+BAO & $0.3174^{+0.0032}_{-0.0036}$ & $0.3172\pm 0.0035$\\
\hline
             & BAO & unconstrained & unconstrained\\
$\epsilon_0$ & SnIa & $-0.060^{+0.090}_{-0.062}$ & $-0.040^{+0.11}_{-0.071}$\\
             & SnIa+BAO & $-0.0008\pm 0.0049$ & $0.0019\pm 0.0061$\\
\hline
             & BAO & --- & unconstrained\\
$\epsilon_1$ & SnIa & --- & $-0.042^{+0.11}_{-0.068}$\\
             & SnIa+BAO & --- & $0.0001\pm 0.0049$\\
\hline
\hline 
\end{tabular}
\end{center}
\end{table}

\begin{figure*}[!t]
	\centering
	\includegraphics[width=0.33\textwidth]{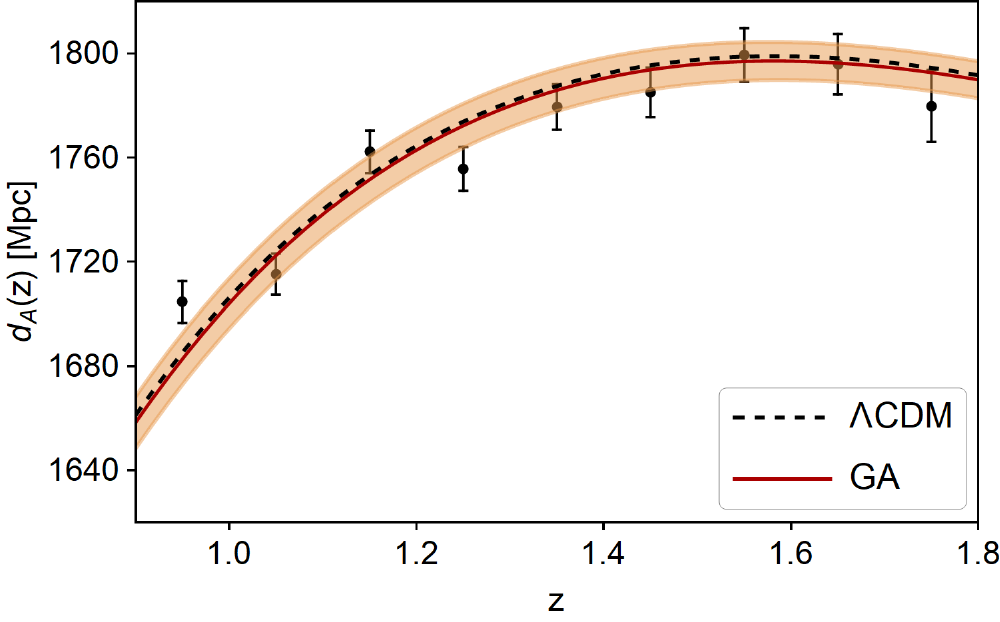}
	\includegraphics[width=0.328\textwidth]{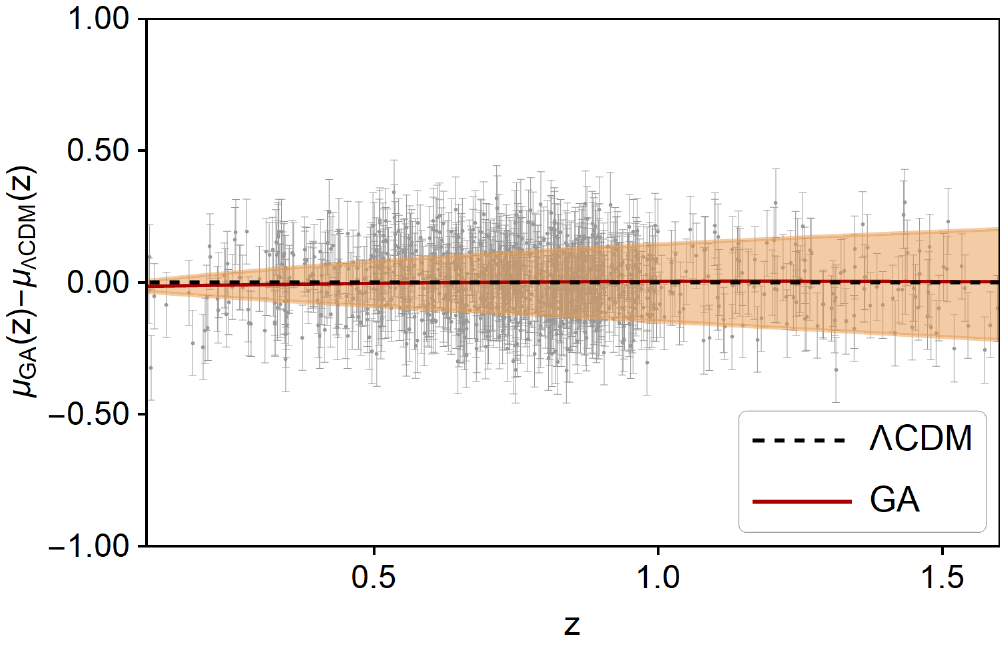}
	\includegraphics[width=0.33\textwidth]{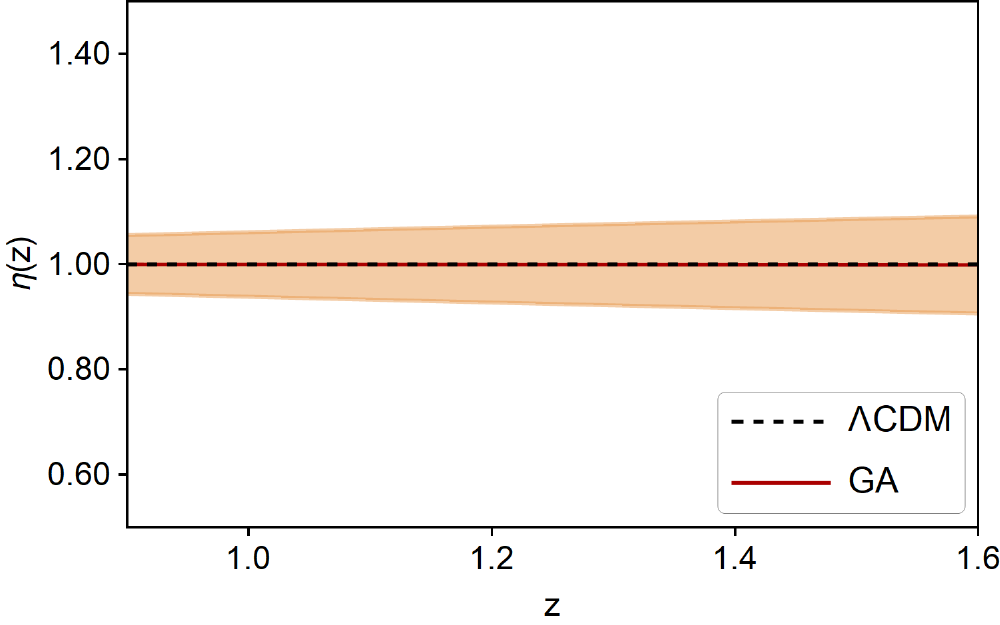}
	\caption{Left: Reconstruction of the angular diameter distance $d_{\rm A}(z)$ from the mock \Euclid data with the GA approach. The dashed black line is $\Lambda$CDM, the solid red line is the GA fit and the shaded region corresponds to the $1\sigma$ GA errors. Centre: Results on the distance modulus $\mu(z)$ from the mock \Euclid+LSST+DESIRE data with the GA approach. The dashed line at zero is $\Lambda$CDM, the red line is the GA fit and the shaded region corresponds to the $1\sigma$ GA errors. For clarity, we only show one thousand of the total SnIa points. Right: Reconstruction of the $\eta(z)$ parameter in the range $z\in[0.9,1.6]$. In all cases the error bars of the data points correspond to $1\sigma$ uncertainty.\label{fig:mockGA}}
\end{figure*}

\begin{figure*}[!t]
	\centering
	\includegraphics[width=0.45\textwidth]{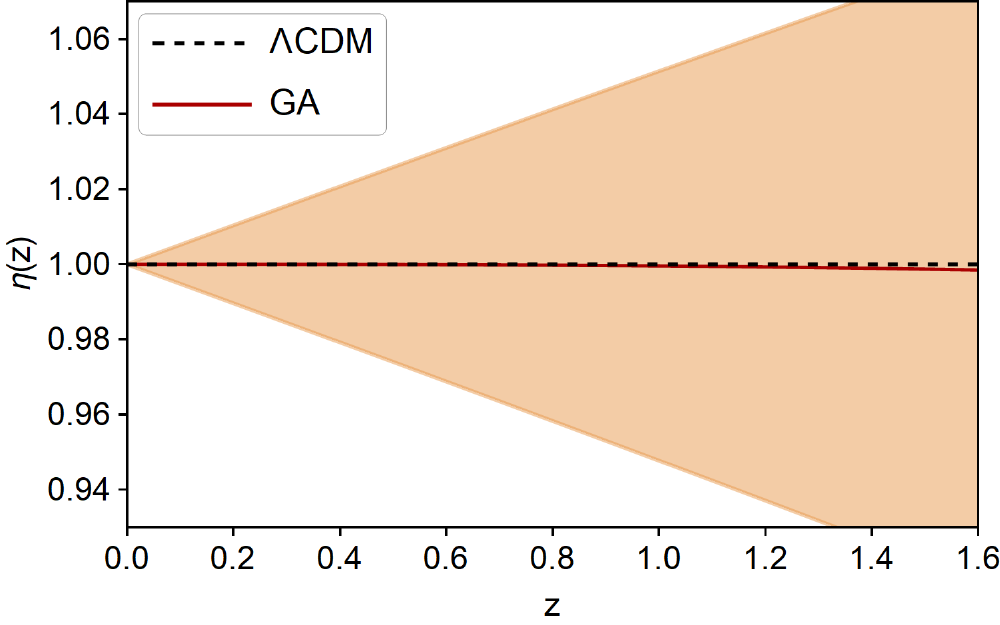}
	\hspace{0.9cm}
	\includegraphics[width=0.45\textwidth]{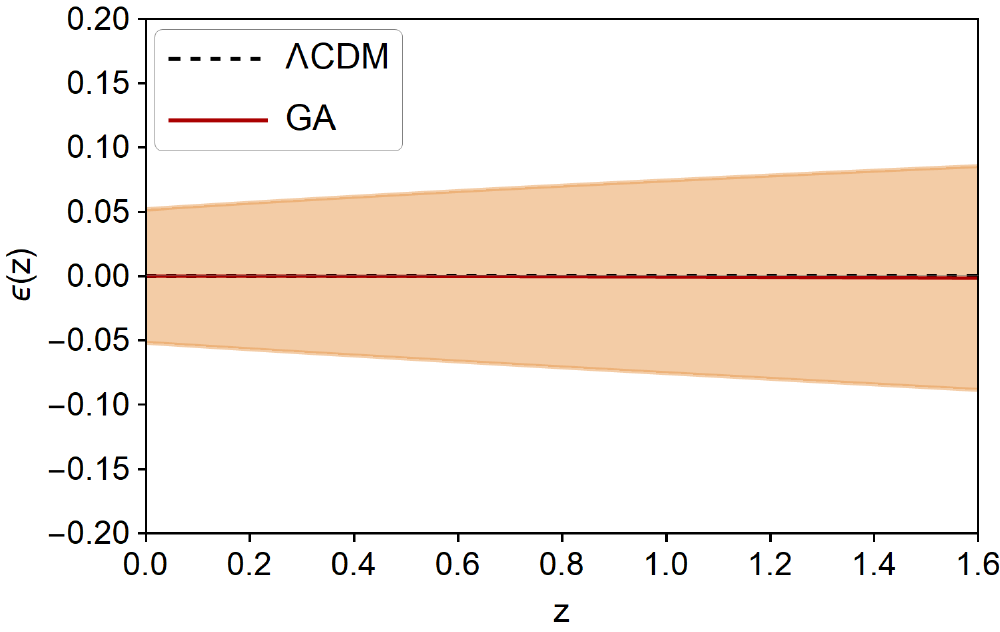}
	\caption{Left: Duality relation $\eta(z)$ for the GA reconstructions for the LSST+DESIRE SnIa data and \Euclid plus DESI BAO mocks. Right: Reconstruction of the $\epsilon(z)$ parameter, calculated via $\epsilon(z)=\frac{\ln{\eta(z)}}{\ln(1+z)}$, is found to be consistent with the fiducial value $\epsilon(z)=0$ within the errors. In both cases the GA reconstruction is the red line, while the shaded region corresponds to the $1\sigma$ errors.\label{fig:mockGAeta}} 
\end{figure*}

As a final analysis, we apply the GAs reconstruction method to the mock BAO and SnIa data. In the left panel of \Cref{fig:mockGA} we show the reconstruction of the angular diameter distance $d_{\rm A}(z)$ with the GA approach using the mock BAO \Euclid data. The dashed black line is the best-fit $\Lambda$CDM model, the solid red line is the GAs fit to the \Euclid BAO data, while the shaded region corresponds to the $1\sigma$ GA errors. In the central panel of \Cref{fig:mockGA} we show the GAs reconstruction of the distance modulus $\mu(z)$ using the LSST+DESIRE simulated data rescaled by the best-fit $\Lambda$CDM, with the latter corresponding to the dashed line at zero. Similarly to the left panel, the red line is the GAs fit and the shaded region corresponds to the $1\sigma$ GAs errors. For clarity, we only show $1000$ out of the total $\sim10\,000$ SnIa points we include in our mock dataset.

Finally, in the right panel of \Cref{fig:mockGA} we show the reconstruction of the $\eta(z)$ function, which is obtained from the joint GA reconstruction of $d_{\rm A}(z)$ and $d_{\rm L}(z)$. It is immediately clear that such a reconstruction is limited if performed using \Euclid BAO data only: The lack of $d_{\rm A}(z)$ measurements for $z<0.9$ forces the method to be applied only in the redshift range $z\in[0.9,1.6]$. Despite this, as can be seen in all panels of \Cref{fig:mockGA}, the GA reconstruction for the mock \Euclid BAO recovers the correct fiducial model in all three cases and in particular, it provides very tight constraints on both the angular diameter distance and the duality parameter $\eta(z)$. Specifically, comparing with the right panel of \Cref{fig:realplotsGA}, we can see that with the GA approach, \Euclid now brings roughly a factor of three improvement compared to the current data with a non-parametric approach.

We also consider the added benefit on the reconstruction brought by DESI BAO data, which cover, as mentioned earlier, the redshift range $z\in[0,0.9]$. In \Cref{fig:mockGAeta} we show the reconstruction of the duality relation $\eta(z)$ (left) and the $\epsilon(z)$ parameter (right) when the full data combination of \Euclid and DESI is considered. We find that, as in the case of the parameterized results, the addition of DESI to the combination of LSST+\Euclid does not improve the constraints  significantly, however in this case we can now cover a wider redshift range. The GA can then recover the fiducial model $\eta(z)=1$ with a $1\%$ error at $z=0.2$ and a $5\%$ error at $z=1$, while the parameter $\epsilon(z)$, obtained inverting \Cref{eq:eta1}, can be measured in a model-independent fashion, with an error between $0.05$ and $0.07$ over the redshift range covered by the data.

\section{Conclusions}\label{sec:conclusions}
In this work, we have constrained deviations from the standard DDR using current and forecast data; for the latter, we focused mainly on the constraints achievable through synergies between \Euclid and contemporary surveys, both for SnIa and BAO. We have discussed in \Cref{sec:theory} several physical mechanisms that can lead to a violation of the DDR, both of astrophysical and BSM origin. For this reason, while we exploited a commonly used parametric approach to constrain the DDR breaking function $\epsilon(z)$, we also used a fully agnostic reconstruction through GAs. The latter approach allows us to obtain constraints without any assumption on the redshift trend of possible deviations from the standard theory.

In the parametric case, within the assumption of a flat $\Lambda$CDM expansion and using the value of $\Omega_{\rm b,0} h^2$ measured by Planck, we found that current SnIa data loosely constrain the deviation from the $\epsilon(z)=0$ limit, due to the degeneracy between the DDR parameters and the total matter energy density  $\Omega_{\rm m,0}$. Such a degeneracy is broken when including BAO data; these are not sensitive to violations of DDR, but tightly constrain $\Omega_{\rm m,0}$, thus yielding tight constraints on the parameterized $\epsilon(z)$. The results obtained with the combination of SnIa and BAO data show that the standard DDR is within $\approx1\sigma$ and the results are compatible with a constant $\epsilon(z)$.

In the case of the machine learning reconstruction, we found that the GA can provide robust constraints in line with the parametric approach, albeit with somewhat larger uncertainties. This is due to the fact that the GA is non-parametric, thus it provides broader and theory-agnostic constraints.  Specifically, we found that in the case of the currently available data, the reconstruction of the duality parameter $\eta(z)$ was fully consistent with the parametric approach and with unity.

Using the same fiducial cosmology assumed in \citetalias{IST:paper1}, we have then created simulated data for upcoming surveys; we focused mainly on the BAO data achievable with \Euclid and on the possible SnIa survey DESIRE that might be provided by this satellite. We complemented the redshift range of \Euclid forecast data with contemporary surveys, namely LSST for SnIa and DESI for BAO.

Analysing these mock data through the parameterized approach, we found an improvement of a factor $\approx6$ with respect to current results when the combination of SnIa and BAO is considered (see \Cref{tab:forecasted_res}). We have also shown in \Cref{fig:forecast_eta} how such constraints translate into a redshift trend for the $\eta(z)$ function, highlighting how the use of mock data significantly improves the bounds on this function, which is now constrained to vary by less than $1\%$ from the fiducial assumption of $\eta(z)=1$. 

Using the GA with the mock data, we reconstructed the duality parameter $\eta(z)$ and we have shown how synergies of \Euclid with other galaxy surveys are crucial in order to be able to use such an approach over an extended range in redshift. With the data combinations considered here we found that the GA can recover the fiducial model $\eta(z)=1$ with an error of $1\%$ at $z=0.2$ and of $5\%$  at $z=1$, as shown in \Cref{fig:mockGAeta}, and with an improvement of roughly a factor of three over the current DDR constraints in the same redshift range. This somewhat less constraining result, compared to the parameterized approach, is mainly due to the completely model-independent and theory-agnostic approach employed here, despite the joint fitting of the SnIa and BAO data.

In summary, our paper highlighted the benefits of synergies between the \Euclid\ BAO survey and external probes in constraining physics beyond the standard model, which could manifest itself through violations of the DDR. In particular, we have demonstrated that such a BAO survey will make it possible to constrain deviations from the DDR at an unprecedented level in the near future using parameterized approaches, while it will also reach a high enough sensitivity to employ model-independent approaches that allow an agnostic reconstruction of possible deviations from the standard DDR. 

\begin{acknowledgements}
We are grateful to P.~Astier for useful discussions and to T.~Baker, G.~Castignani and P. Ntelis for comments on the manuscript.\\

MM has received the support of a fellowship from ``la Caixa” Foundation (ID 100010434), with fellowship code LCF/BQ/PI19/11690015, and the support of the Spanish Agencia Estatal de Investigacion through the grant “IFT Centro de Excelencia Severo Ochoa SEV-2016-0597”. The work of CJM was financed by FEDER---Fundo Europeu de Desenvolvimento Regional funds through the COMPETE 2020---Operational Programme for Competitiveness and Internationalisation (POCI), and by Portuguese funds through FCT - Funda\c c\~ao para a Ci\^encia e a Tecnologia in the framework of the project POCI-01-0145-FEDER-028987. SN acknowledges support from the research project  PGC2018-094773-B-C32, the Centro de Excelencia Severo Ochoa Program SEV-2016-059 and the Ram\'{o}n y Cajal program through Grant No. RYC-2014-15843.  DS acknowledges financial support from the Fondecyt Regular project number 1200171. IT acknowledges support from the Spanish Ministry of Science, Innovation and Universities through grant ESP2017-89838-C3-1-R, and the H2020 programme of the European Commission through grant 776247. AA acknowledges support from the Science and Technology Facilities Council (STFC) grant ST/P000703/1. VY acknowledges funding from the European Research Council (ERC) under the European Union’s Horizon 2020 research and innovation programme (grant agreement No. 769130)\\

\AckEC
\end{acknowledgements}

\begin{appendix} 

\section{The current BAO data}\label{app:BAOchi}
Here we describe the currently available BAO data we use in our analysis. In particular, we use the measurements from 6dFGS \citep{Beutler:2011hx}, SDDS \citep{Anderson:2013zyy}, BOSS CMASS \citep{Xu:2012hg}, WiggleZ \citep{Blake:2012pj}, MGS \citep{Ross:2014qpa} and BOSS DR12 \citep{Gil-Marin:2015nqa}, DES \citep{Abbott:2017wcz}, Lya \citep{Blomqvist:2019rah}, DR14 LRG \citep{Bautista:2017wwp} and quasars \citep{Ata:2017dya}.

The data provided by these surveys are described by the function $d_z$, defined in \Cref{sec:curr_data}. The 6dFGs  and WiggleZ BAO data are 
\be
\begin{array}{ccc}
 z  & d_z & \sigma_{d_z } \\
 \hline
 0.106 & 0.336 & 0.015 \\
 0.44 & 0.073 & 0.031 \\
 0.6 & 0.0726 & 0.0164 \\
 0.73 & 0.0592 & 0.0185 \\
\end{array}\, ,
\ee
with their inverse covariance matrix given by
\be C_{ij}^{-1}=\left(
\begin{array}{cccc}
 4444.4 & 0 & 0 & 0 \\
 0 & 1040.3 & -807.5 & 336.8 \\
 0 & -807.5 & 3720.3 & -1551.9 \\
 0 & 336.8 & -1551.9 & 2914.9 \\
\end{array}
\right)\, ,\ee
and with the $\chi^2$ being
\be 
\chi^2_{\rm 6dFS,Wig}=V^i C_{ij}^{-1} V^j,
\ee
and the data vector $V^i=d_{z,i}-d_z(z_i,\Omega_{\rm m, 0})$.

The BAO measurements from MGS and SDSS (LOWZ and CMASS samples) are given by $D_V/r_{\rm s} = 1/d_z$ via 
\be\begin{array}{ccc}
 z  & 1/d_z & \sigma_{1/d_z } \\
 \hline
 0.15 & 4.46567 & 0.168135 \\
 0.32 & 8.62 & 0.15 \\
 0.57 & 13.7 & 0.12 \\
\end{array}\, ,\ee
and the $\chi^2$ is then
\be 
\chi^2_{\rm MGS,SDSS}=\sum_i \left(\frac{1/d_{z,i}-1/d_z(z_i,\Omega_{\rm m, 0})}{\sigma_{1/d_{z,i}}}\right)^2.
\ee

The BAO data from DES is of the form $d_{\rm A}(z)/r_{\rm s}$ with the data vector $(z,d_{\rm A}(z)/r_{\rm s},\sigma)=(0.81, 10.75, 0.43)$ and the $\chi^2$ being
\be 
\chi^2_{\rm DES}=\sum_i \left(\frac{d_{\rm A}({z,i})/r_{\rm s}-d_{\rm A}(z_i,\Omega_{\rm m, 0})/r_{\rm s}}{\sigma_{d_{\rm A}(z,i)/r_{\rm s}}}\right)^2.
\ee

The BAO data from Ly-$\alpha$ are of the form $f_{\rm BAO}=\left((1+z)\;d_{\rm A}/r_{\rm s}, D_{\rm H}/r_{\rm s}\right)$ and are given by
\be
\begin{array}{ccccc}
 z & (1+z)\;d_{\rm A}/r_{\rm s}  &  \sigma_{(1+z)\;d_{\rm A}/r_{\rm s}} & D_{\rm H}/r_{\rm s}  &  \sigma_{D_{\rm H}/r_{\rm s}}\\
 \hline
 2.35 & 36.3 & 1.8 & 9.2 & 0.36\\

\end{array}\, ,
\ee 
with the $\chi^2$ being
\be 
\chi^2_{\text{Ly-}\alpha}=\sum_i \left(\frac{f_{\text{BAO},i}-f_{\rm BAO}(z_i,\Omega_{\rm m, 0})}{\sigma_{f_{\rm BAO}}}\right)^2.
\ee

The DR14 LRG and quasar BAO data assume $r_{\rm s,fid} = 147.78$ and are given by $D_V/r_{\rm s} = 1/d_z$ 
\be
\begin{array}{ccc}
 z  & 1/d_z & \sigma_{1/d_z } \\
 \hline
 0.72 & 2353/r_{\rm s,fid} & 62/r_{\rm s,fid} \\
 1.52 & 3843/r_{\rm s,fid} & 147/r_{\rm s,fid} \\
\end{array}\, ,
\ee 
and the $\chi^2$ being
\be 
\chi^2_{\rm LRG,Q}=\sum_i \left(\frac{1/d_{z,i}-1/d_z(z_i,\Omega_{\rm m, 0})}{\sigma_{1/d_{z,i}}}\right)^2.
\ee
Finally, the total $\chi^2$ is 
\be
\chi^2_{\rm tot}=\chi^2_{\rm 6dFS,Wig}+ \chi^2_{\rm MGS,SDSS} + \chi^2_{\rm DES} +\chi^2_{\text{Ly-}\alpha}+\chi^2_{\rm LRG,Q}.
\ee

\end{appendix}

\bibliographystyle{aa} 
\bibliography{references} 

\end{document}